\documentclass[preprint]{elsarticle}
\usepackage{silence}
\usepackage{graphicx}
\usepackage{epstopdf}
\usepackage{lineno,hyperref}
\usepackage{multirow}
\usepackage{amsmath}
\usepackage{caption} 
\usepackage{booktabs}
\usepackage{longtable}
\usepackage{diagbox} 
\usepackage{multirow}
\usepackage{color}
\usepackage{url}
\usepackage{array}
\usepackage{siunitx}
\modulolinenumbers[5]
\graphicspath{{figures/}}
\usepackage{color}
\usepackage{xcolor}
\usepackage{comment}
\usepackage{lineno} 

\journal{Nuclear Instruments and Methods in Physics Research Section A}
 
\begin{document}
\modulolinenumbers[1]

\begin{frontmatter}
\title{The High Level Trigger and Express Data Production at STAR \\ }
\author[1]{Wayne Betts}
\author[2]{Jinhui Chen}
\author[1]{Yuri Fisyak}
\author[1]{Hongwei Ke}
\author[3,4,5,6]{Ivan Kisel\corref{cor1}}
\ead{I.Kisel@compeng.uni-frankfurt.de}
\author[3]{Pavel Kisel}
\author[3]{Grigory Kozlov}
\author[1]{Jeffery Landgraf}
\author[1]{Jerome Lauret}
\author[7]{Tonko Ljubicic}
\author[2,8]{Yugang Ma}
\author[9]{Spyridon Margetis}
\author[10]{Hao Qiu}
\author[10]{Diyu Shen}
\author[2]{Qiye Shou}
\author[11]{Xiangming Sun}
\author[1]{Aihong Tang\corref{cor1}}
\ead{aihong@bnl.gov}
\author[1]{Gene Van Buren}
\author[6]{Iouri Vassiliev}
\author[2]{Baoshan Xi}
\author[12]{Zhenyu Ye}
\author[1]{Zhengqiao Zhang}
\author[4]{Maksym Zyzak}

\affiliation[1]{organization={Brookhaven National Laboratory}, addressline={PO Box 5000}, city={Upton}, postcode={NY 11973}, country={USA}}

\affiliation[2]{organization={Key Laboratory of Nuclear Physics and Ion-beam Application (MOE), Institute of Modern Physics, Fudan University}, addressline={220 Handan Rd.}, city={Shanghai}, postcode={200433}, country={China}}

\affiliation[3]{organization={Goethe University}, addressline={Theodor-W.-Adorno-Platz 1}, city={Frankfurt am Main}, postcode={60323}, country={Germany}}

\affiliation[4]{organization={FIAS, Frankfurt Institute for Advanced Studies}, addressline={Ruth-Moufang-Str.\ 1}, city={Frankfurt am Main}, postcode={60438}, country={Germany}}

\affiliation[5]{organization={HFHF, Helmholtz Research Academy Hesse}, addressline={Max-von-Laue-Str.\ 12}, city={Frankfurt am Main}, postcode={60438}, country={Germany}}

\affiliation[6]{organization={GSI, Helmholtz Center for Heavy Ion Research}, addressline={Planckstr.\ 1}, city={Darmstadt}, postcode={64291}, country={Germany}}

\affiliation[7]{organization={Rice University}, addressline={6100 Main St.}, city={Houston}, postcode={TX 77005}, country={USA}}

\affiliation[8]{organization={School of Physics, East China Normal University,}, addressline={ Physics Building, Minhang District.}, city={Shanghai}, postcode={200241}, country={China}}

\affiliation[9]{organization={Kent State University}, addressline={800 E Summit St.}, city={Kent}, postcode={OH 44240}, country={USA}}

\affiliation[10]{organization={Institute of Modern Physics, Chinese Academy of Sciences}, addressline={509 Nanchang Rd.}, city={Lanzhou}, postcode={730000}, country={China}}

\affiliation[11]{organization={Central China Normal University}, addressline={152 Luoyu Avenue}, city={Wuhan}, postcode={430079}, country={China}}

\affiliation[12]{organization={University of Illinois at Chicago}, addressline={1200 W Harrison St.}, city={Chicago}, postcode={IL 60607}, country={USA}}

\cortext[cor1]{Corresponding author}

\begin{abstract}

To meet the demands of the Beam Energy Scan phase-II (BES-II) program, the STAR experiment at the Relativistic Heavy Ion Collider (RHIC) developed a dual real-time framework consisting of a High Level Trigger (HLT) and an Express Data Production system (xProduction). 

The HLT operates online within the Data Acquisition (DAQ) chain on a dedicated multicore CPU cluster with the option to offload compute-intensive kernels to Xeon Phi coprocessors. It uses parallelized algorithms, such as the Cellular Automaton (CA) Track Finder, to perform rapid tracking, vertexing, and event filtering. This allows it to select events of interest in real time and provide immediate feedback on detector and beam conditions.

In contrast, the xProduction workflow runs concurrently and independently of the DAQ loop. It applies near offline-quality calibration and reconstruction within hours of data collection. The xProduction input is the express data stream, whose content can be enriched by HLT trigger/priority selections under DAQ/HLT resource constraints, and it uses the STAR calibration/conditions framework, incorporating online calibration/QA information when available. This enables early preliminary physics analysis, including the reconstruction of rare signals, such as hyperons and hypernuclei. It also provides collaboration-wide access to analysis-ready datasets.

Together, the HLT and xProduction systems form a complementary architecture: the HLT performs online event selection while the xProduction chain delivers high-quality results in a short amount of time. This integrated framework has enabled the prompt reconstruction of the \texorpdfstring{${}^5_{\Lambda}\mathrm{He}$}{5-Lambda-He} hypernucleus with high statistical significance and the processing of hundreds of millions of heavy-ion collision events efficiently. Its demonstrated scalability and robustness establish a model for future high-luminosity experiments requiring both online event filtering and rapid access to analysis-quality data.

\end{abstract}

\end{frontmatter}

\section{Introduction}

One of the central goals of modern nuclear physics is the study of strongly interacting matter under extreme conditions. The STAR experiment at RHIC investigates the properties of nuclear matter at high temperatures and densities, where phenomena such as the quark-gluon plasma~\cite{STAR:2005gfr, PHENIX:2004vcz, BRAHMS:2004adc, PHOBOS:2004zne} and exotic states like hypernuclei~\cite{STAR:2010gyg, STAR:2023fbc} can be studied. The Beam Energy Scan programs, BES-I~\cite{STAR:2010vob, Chen:2024aom} and BES-II~\cite{Aparin2023}, focus on low-energy collisions to probe the Quantum Chromodynamics (QCD) phase diagram at high net baryon density. Compared to BES-I, BES-II combined detector/readout upgrades and higher delivered statistics, resulting in substantially larger datasets and increased demand for prompt reconstruction, calibration, and data-quality feedback. Such capabilities are broadly beneficial for precision heavy-ion measurements and related experimental and phenomenological studies (see, e.g., Refs.~\cite{Jia:2022ozr, Giacalone:2024bud, STAR:2022fan, Huang:2024baryonBQ, Ma:2026quarkSoupTemp}, among many others).

Realizing these objectives, especially given the complex final states analyzed in BES-II and the
need to capture rare signals, requires prompt turnaround from data taking to analysis such that sufficient quality data can be assured. However, timely turnaround is constrained by large data volumes and the inherent latency
of the multi-stage offline workflow (calibration, reconstruction, and quality assurance analyses). By default,
STAR follows a traditional offline workflow: raw DAQ data are recorded and archived to
the High Performance Storage System (HPSS). After the data-taking period (typically the
annual run) concludes, the data are staged from HPSS and reconstructed on the central
farm in large production passes. Calibration jobs and dedicated runs provide updated detector calibrations, followed by full reconstruction and Quality Assurance (QA);
analysis-ready datasets become available only after these passes complete, introducing
latencies of months to years and limiting immediate operational feedback. STAR has also operated a "FastOffline" reconstruction at the central Scientific Computing and Data Facilities (SCDF) at BNL since first collisions, with hour-scale latency for a smaller fraction ($\sim5\%$) of the incoming data. Its use of shared resources and more strictly controlled software have served many calibration and quality assurance purposes very well, but limited experimentation for fast physics-level reconstruction (e.g. hyperon/hypernuclei yield checks).

To provide a concrete scale for STAR operations, from 2009-2022, the typical output rate for min-bias AuAu at about 40\% dead time was roughly 1.8~kHz, with the dead time dominated by DAQ/readout busy time (event size and output/buffer limitations),  although in practice operation was often configured to run closer to $\sim$1 kHz to optimize sampled luminosity. At that lower setpoint, high-multiplicity events can be recorded at about 15\% dead time. For the run 2023 and later, these limits were raised: the system supported $\sim$4 kHz minimum-bias at 40\% dead time, and $\sim$2 kHz high-multiplicity at 15\% dead time. Here, minimum-bias denotes a very inclusive trigger intended to record a broadly representative sample of collisions across impact parameters with minimal selection bias. The instantaneous input collision rate can be much higher than the recorded DAQ rate: up to $\sim$100 kHz for Au+Au and $\sim$600 kHz for p+p. During minimum-bias running,  RHIC is often requested by experiments to deliver a reduced collision rate of order 10 kHz to control pile-up while still saturating available bandwidth.  Rates vary with species and energy; in particular, low-energy Au+Au runs operate at lower collision rates. For the BES-II scan, input rates were typically 150-1500~Hz.

To overcome this latency, STAR implemented a dual real-time framework consisting of a High Level Trigger (HLT) and an xProduction chain. The HLT runs on a dedicated multi-core CPU cluster online with optional Xeon Phi acceleration. It performs fast tracking, vertexing, and event filtering directly within the DAQ stream. This provides prompt data-quality feedback and enables initial reduction by selecting enriched event samples. In parallel, the xProduction system runs on the same HLT cluster during data acquisition. It processes the HLT-selected events that are written in DAQ format to the HLT distributed storage (before archival to HPSS) and, using the same calibration snapshots as the offline chain, performs near offline-quality reconstruction within hours and produces compact picoDst (compact, ROOT-tree analysis ntuples) datasets at hundreds of Hz. These outputs are approximately two orders of magnitude smaller than the raw data yet still preserve the necessary physics content for immediate analysis.

This complementary approach, using the HLT for online selection and the xProduction system for rapid, high-quality reconstruction, provides efficient data reduction and a fast turnaround time. This allows for the monitoring of rare physics signals in real time while maintaining full archival storage of the raw stream. Thus, it bridges the gap between immediate operational needs and long-term offline processing, supporting BES-II’s goal of taking precise measurements of rare hyperons, hypernuclei, and other exotic probes~\cite{71b1995ba77349bb8ba5bf9b71c53489, ALICE:2018phe, CMSTrigger:2005yhe, Aaij:2019zbu, Leung:2023gki, STAR:2021orx, Chen:2023mel}.

\section{The High Level Trigger}

STAR employs a pipelined, multi-level trigger~\cite{Bieser:2002ah}. Level-0 is implemented in custom VME hardware: a tree of Data Storage and Manipulation (DSM) boards combines fast detector inputs each RHIC clock tick; the Trigger Control Unit (TCU) issues the trigger command and token, and the Trigger Clock Distribution fans out per-detector actions under the RHIC clock distributed by the RHIC Clock and Control unit.  Software Levels-1 and Level-2 run on the CPUs that control the TCU and the event builders, respectively.  These software levels are used for tasks such as pile-up suppression and trigger monitoring, but they do not have access to the full event data nor to substantial computing resources.  Here "pile-up" refers to multiple collisions occurring within the same detector readout window and being recorded as a single overlaid event.

 The STAR HLT constitutes the final software stage - level-4 (L4)--of the multi-stage trigger chain~\cite{Bieser:2002ah, Landgraf:2002zw}, progressively narrowing down events from high-rate, simple triggers to data-intensive, complex analyses. In STAR, “L3” referred to an earlier online display/monitoring mechanism and is now obsolete; the term L4 is used here to avoid ambiguity with that historical usage. Unlike lower level triggers that apply straightforward event filtering, the HLT accesses detailed data from multiple detectors, performing sophisticated event reconstruction and filtering with high granularity. The precision vertex provided by the HLT was critical for RHIC beam tuning at low energies, because the RHIC beam position monitors and STAR low-level trigger detectors could not reliably distinguish true beam-beam collisions from beam-beampipe interactions.

\subsection{HLT Computer Farm}

During BES-II, STAR’s HLT farm comprised 27 Linux nodes. Eighteen nodes had more than 40 CPU cores; the remainder had fewer. Twenty-two nodes housed two Intel Xeon Phi 7110P coprocessors and one housed a single card, for a total of 45. In aggregate the farm provided 1,192 logical CPU cores. About 200 cores were allocated to real-time processing, with the balance supporting express data calibration and production, on a configuration tuned for high-throughput online reconstruction and calibration.  This system processes a subset of events into picoDst files in near real-time, significantly reducing the lag associated with conventional data transfer and processing. The xProduction (discussed in detail later) capitalizes on HLT's substantial computing resources, running jobs in an HTCondor-managed queue. These nodes perform rapid data processing directly on the HLT's distributed filesystem (Ceph; see Fig.~\ref{fig-STAR-xHLT}) which offers 120~TB of usable storage with robust read/write performance.

\subsection{Software Structure and Key Components}

The HLT’s event reconstruction workflow mirrors offline analysis processes, but is adapted for real-time operation, transforming raw detector data into physical quantities such as tracks and vertices. The system’s primary functions include event reconstruction and event filtering using physics-driven criteria, for example, a Heavy Fragment trigger that enriches hypernuclei searches by selecting events with large ionization ($dE/dx$) tracks, and diElectron and diMuon triggers that retain events with $e^+e^-$ and $\mu^+\mu^-$ pairs for $J/\psi$ and heavy-flavor studies.

Within the STAR detector layout, a large-acceptance solenoidal spectrometer with a central Time Projection Chamber (TPC) surrounded by Time-of-Flight (TOF) and other subsystems (see the detector overview in ~\cite{STAR:2002eio}), the HLT software is designed with a modular structure. Data from each detector are first processed independently to produce detector-specific information. This is followed by a multi-detector matching phase, such as correlating TPC tracks~\cite{Anderson:2003ur} with TOF hits~\cite{Llope:2012zz}, to assemble complete event information. In the final stage, selection algorithms evaluate each event against predefined physics criteria, retaining those that meet the necessary requirements.

Online tracking exploits STAR’s sectorized TPC geometry. CA
seeding forms short segments from hit triplets across adjacent pad rows within a
sector, with dedicated treatment at sector boundaries; seeds are then extended by
a Kalman filter and merged across sectors to produce global tracks. During running,
rolling updates (see section ~\ref{subsec:HLTCalibration}) to space-charge and beamline parameters are applied before and during
the fit, and a primary-vertex–constrained refit is used where appropriate to obtain
primary track parameters. For subsystem association, TPC tracks are projected to TOF~\cite{Llope:2012zz} and Barrel Electromagnetic Calorimeter (BEMC)~\cite{STAR:2002ymp}, and matched to nearby signals using straightforward geometric and timing-consistency checks, providing
time-of-flight and calorimetric information for lightweight Particle Identification (PID) and QA. These associations,
built with the current HLT calibrations, feed the run-by-run QA plots and populate the
xProduction picoDsts.

\begin{figure}[htbp]
\centering
\includegraphics[width=0.9\textwidth]{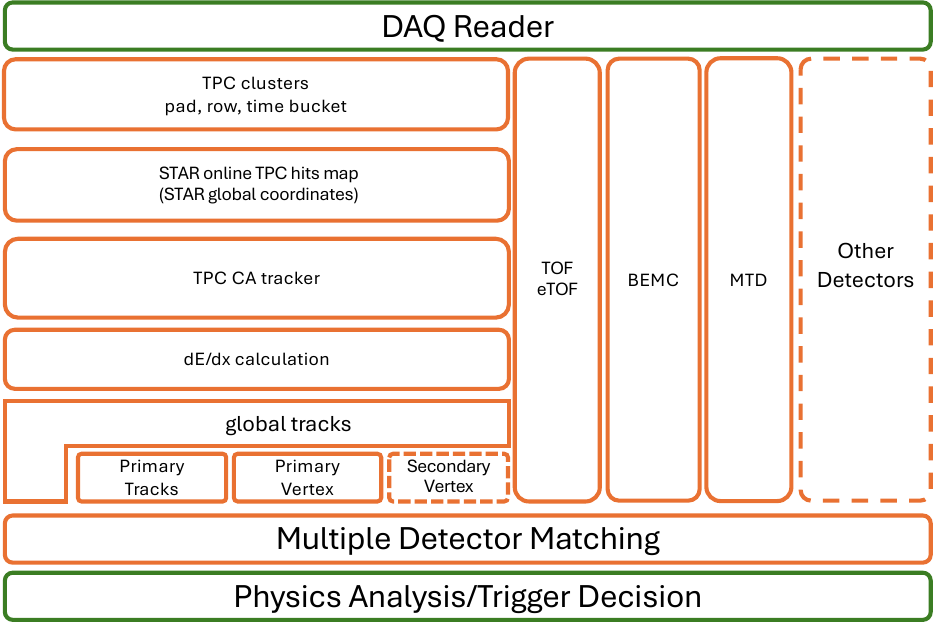}
\caption{
HLT event reconstruction pipeline. Information from sub-detectors, including TPC~\cite{Anderson:2003ur}, TOF~\cite{Llope:2012zz}, BEMC~\cite{STAR:2002ymp}, and Muon Telescope Detector (MTD)~\cite{Yang:2014xta} etc., are processed independently then jointly assembled to form a complete event to be used in trigger decision making.}
\label{fig:HLT_pipleline}
\end{figure}

This modular architecture and parallel processing framework provide flexibility, allowing the HLT to support multiple event analyses simultaneously. The adaptable design allows for the continuous refinement of selection algorithms and the incorporation of new analyses as experimental priorities evolve. Together, these features enable the HLT to deliver high-quality data in real time while interfacing tightly with STAR’s DAQ and detector-control systems~\cite{Landgraf:2002zw}. Fig.~\ref{fig:HLT_pipleline} presents the HLT event reconstruction pipeline and key components.

\subsection{Integration with STAR Data Acquisition}\label{subsec:DAQIntegration}

\begin{figure}[h!]
	\centering
	\includegraphics[width=12.cm]{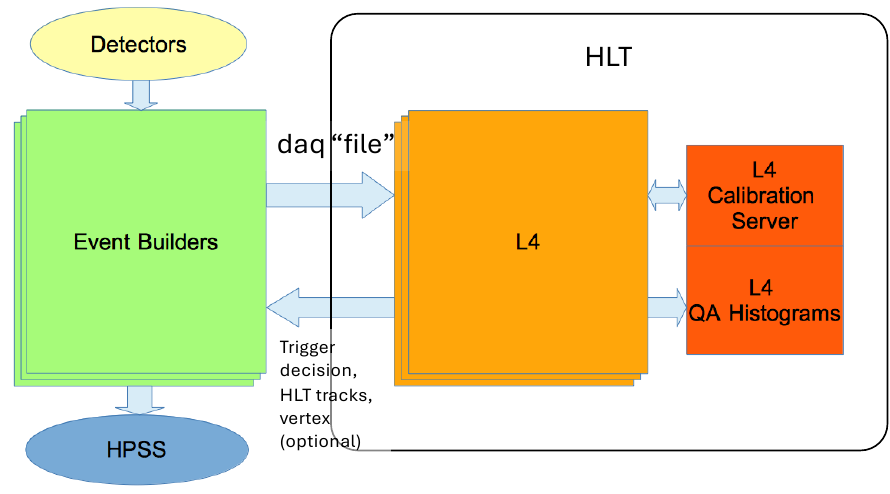}\hspace{0.5cm}
	\includegraphics[width=12.cm]{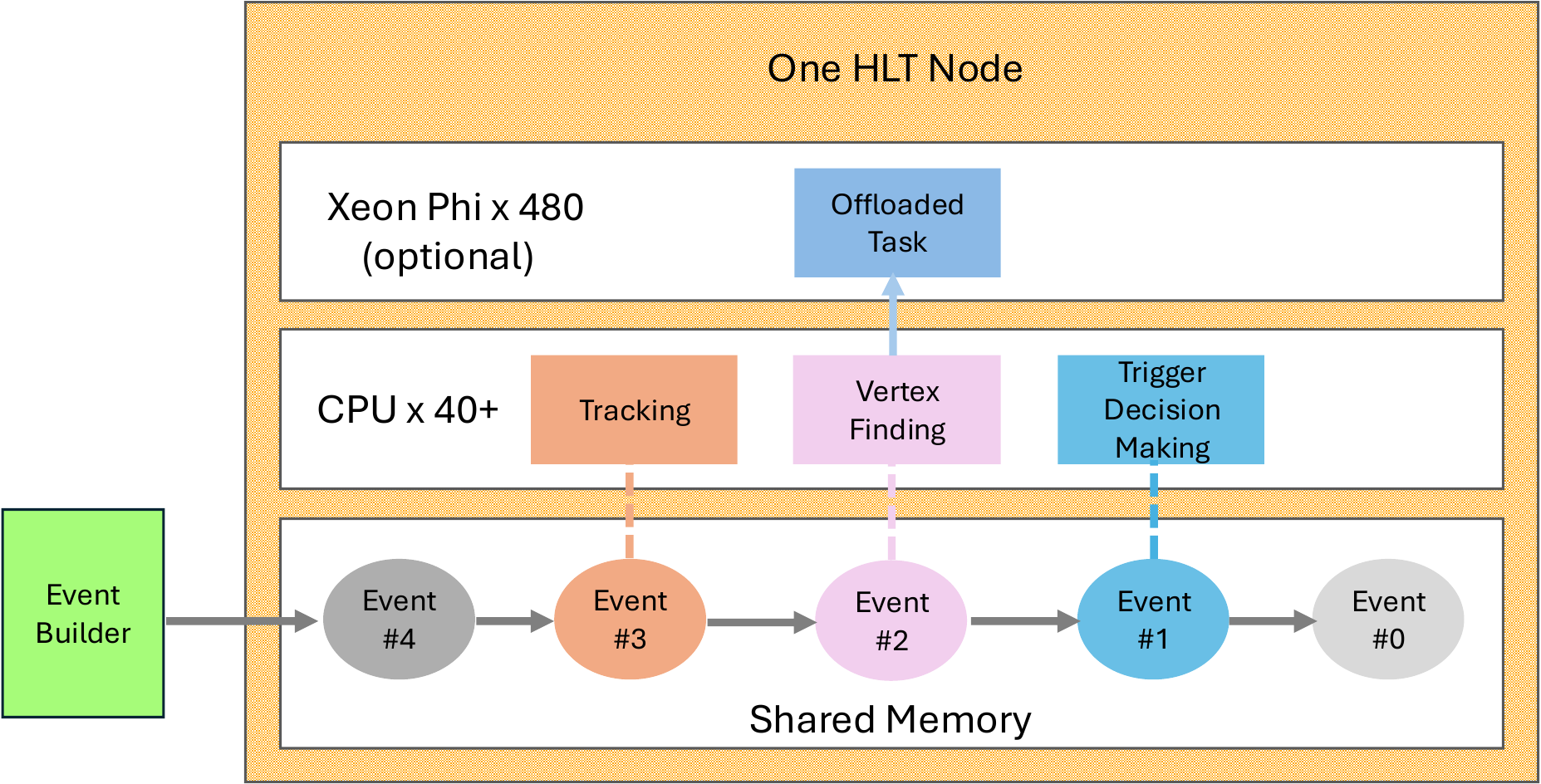}    
	\caption{Top: Scheme of HLT integration with Data Acquisition system at STAR. 
    Bottom: Workload distribution in one HLT node. Because rates vary with beam energy and trigger mix, an illustrative benchmark is included : for Au+Au collisions at $\sqrt{s_{\rm NN}}$= 200 GeV the HLT sustained $\sim$ 2 kHz processing ($\sim$ 1.7-2 GB/s). BES-II input rates are lower; 200 GeV is shown as an upper-envelope stress case. Trigger-specific rates depend on the selection; e.g., the diMuon trigger had $\sim$ 1$\%$ acceptance.
    } 
	\label{fig:HLT_DAQ-oneHLTNode}
\end{figure}

The HLT system is deeply integrated with STAR's DAQ~\cite{Landgraf:2002zw} to ensure efficient and responsive event handling during extended experimental runs, while all raw events are recorded and archived to HPSS; the
HLT/xProduction path provides real-time reduction for prompt QA and early
physics. On each HLT node, a resident service runs continuously in an event-driven mode: it listens for DAQ run-control commands (e.g., run start/stop), performs the required configuration and data processing when a command arrives, and then returns to the listening state. Fig.~\ref{fig:HLT_DAQ-oneHLTNode} (top) summarizes the HLT–DAQ integration. Using parallel processing, multiple HLT tracking nodes coordinate data flow and generate trigger decisions, allowing transition between data taking and online processing to follow the standard DAQ sequence without additional HLT-imposed deadtime.

 The system uses stage-bounded synchronization: worker threads are independently scheduled and synchronize only at the boundaries between reconstruction stages and at the final decision step; intermediate
products remain in node-local shared memory and are handed off between
stages (no cross-node shared state). This keeps all stages busy and exposes parallelism while distributing work efficiently across CPUs and coprocessors. Scalability is built in; within a node, reconstruction is split into stage-specific tasks that workers pull from queues; across nodes, events are partitioned among multiple tracking nodes under run control, so capacity can be increased by enabling additional nodes without changing the algorithms. With this setup, the HLT sustains $\sim$2~kHz processing for minimum bias Au+Au collisions at $\sqrt{s_{\mathrm{NN}}}=200~\mathrm{GeV}$, which matched and exceeded STAR’s DAQ-delivered rates during 200~GeV collider running (typically $\sim$1.5–1.7~kHz at $\sim$1.7~GB/s) and thus met operational requirements. This capacity also comfortably covered BES-II rates (150–1500~Hz). STAR subsequently upgraded the TPC readout to a 5~kHz DAQ (“DAQ5k”) in Run~23 (2023). Following this upgrade, the HLT has typically been operated on a prescaled (downsampled) fraction of the DAQ-delivered events primarily for online monitoring and QA rather than processing the full DAQ stream; in this mode the locally buffered HLT-selected stream is correspondingly prescaled, while the full raw DAQ data remain archived for offline reconstruction, as full-stream 5~kHz processing would require additional online compute beyond the deployed farm.

\subsection{Parallelization and Xeon Phi offload}\label{subsec:PhiOffload}
Parallelism is exploited at two levels (Fig.~\ref{fig:HLT_DAQ-oneHLTNode}, bottom). At the farm level, independent events are distributed across HLT nodes, so throughput scales with the number of active nodes. Within each node, threads execute a multistage pipeline, with online tracking and vertex finding feeding the final L4 trigger decision, so that multiple events are in flight concurrently while the decider serializes the output.

Compute-intensive short-lived–particle reconstruction (KF Particle Finder~\cite{MZyzak, PKisel}, sec. ~\ref{subsec:KFParticle}) is vectorized and offloaded to a Xeon~Phi coprocessor with event-level parallelism. Tasks that are I/O-bound or require global context, such as Cellular Automaton (sec.~\ref{subsec:CATrackFinder}) tracking, primary-vertex finding, detector matching, and the L4 decision, remain on the host CPUs. If a coprocessor is unavailable, the same code path executes on the CPUs (vectorized), preserving algorithmic equivalence. This division of labor keeps the L4 decision path CPU-resident and non-blocking with respect to the coprocessor stage, which stabilizes end-to-end latency. When KFParticle reconstruction is enabled, offloading reduces host-CPU contention and typically increases sustainable per-node throughput at the lower tens-of-percent level. 

\subsection{Real-Time Calibration}\label{subsec:HLTCalibration}

Accurate calibration is essential to the HLT system, ensuring precision in real-time event reconstruction across STAR’s detectors.
Calibration distribution is coordinated by the HLT control server, which
hosts the central calibration and configuration repository; an auto-calibration
service running on the control server ingests lightweight summaries from the HLT
nodes during data taking and periodically publishes updated constants. Tracking
nodes subscribe to these updates and replace the corresponding in-memory tables,
so subsequent events use the new snapshot; the same snapshot is mirrored to the
central repository for provenance and restart consistency. 
The HLT calibration approach prioritizes low-latency data access by using pre-generated calibration files stored locally on each tracking node, reducing dependency on networked storage or databases. These files are periodically updated through a centralized git repository, hosted on the HLT control server; each tracking node maintains
a local clone and automatically pulls updates at regular intervals (e.g., every
15~min) to stay synchronized during running. 

Calibration data for detectors like the TPC, TOF, and BEMC are incorporated into
HLT processing workflows. The TPC calibration, vital for precise track reconstruction,
includes corrections for space-charge and alignment. Space-charge distortions are corrected via per-sector
TPC hit maps generated from recent DAQ files with the fast offline chain, using
parameters from the offline database, and scaled using live coincidence rate from either Zero Degree Calorimeter~\cite{Adler:2000bd} or Beam-Beam Counter~\cite{Whitten:2008zz} as a luminosity proxy for the instantaneous space-charge load; Alignment constants are taken from the offline
database and distributed with the hit maps, while the beamline position is refined
online by the auto-calibration service. In addition, the HLT’s auto-calibration
service dynamically adjusts key parameters (e.g., TPC d$E$/d$x$ gain
equalization and beamline position) via rolling updates and broadcasting to all HLT nodes. This feedback loop allows the HLT to account for fluctuating conditions over
extended runs, enhancing overall tracking accuracy and stability; operationally, the mean of distance of closest approach to beam line in the transverse plane,
$\langle DCA_{xy}\rangle$, is maintained within $\sim$0.01\,cm of zero over long runs, while
space-charge and beamline parameters are refreshed every
$\mathcal{O}(10^2\!-\!10^3)$ events (seconds at $\sim$1.5\,kHz). The service with
rolling updates of TPC d$E$/d$x$ gain equalization and the beamline estimate based on
per-node summaries meets the accuracy required for online tracking, vertexing,
and trigger selections, with QA showing agreement to offline vertices and PID bands
(see Fig.~\ref{fig:Vxy_L4_offLine},~\ref{fig:dEdx_P-TOF_inverseBeta_P}).

\subsection{Quality Assurance and Event Visualization}

The HLT system’s QA and event visualization tools play a critical role in ensuring data integrity and enabling real-time diagnostics. HLT QA monitors physics-driven parameters, such as track momentum, primary vertex location, and reconstructed particle masses, allowing for immediate detection of issues that may arise from detector misalignments, calibration drift, or unexpected signal behavior.

During each run, QA metrics are continuously updated and accessible to operators, providing a real-time overview of data quality. These metrics enable swift identification of anomalies, while at the conclusion of each run, a detailed QA summary is saved for documentation and further analysis. For more thorough assessments, the HLT compiles daily summaries and aggregated QA plots, enabling longitudinal studies that identify trends or recurring issues and inform ongoing optimization of detector performance.

\begin{figure}[htbp]
\centering
\includegraphics[width=1.0\textwidth]{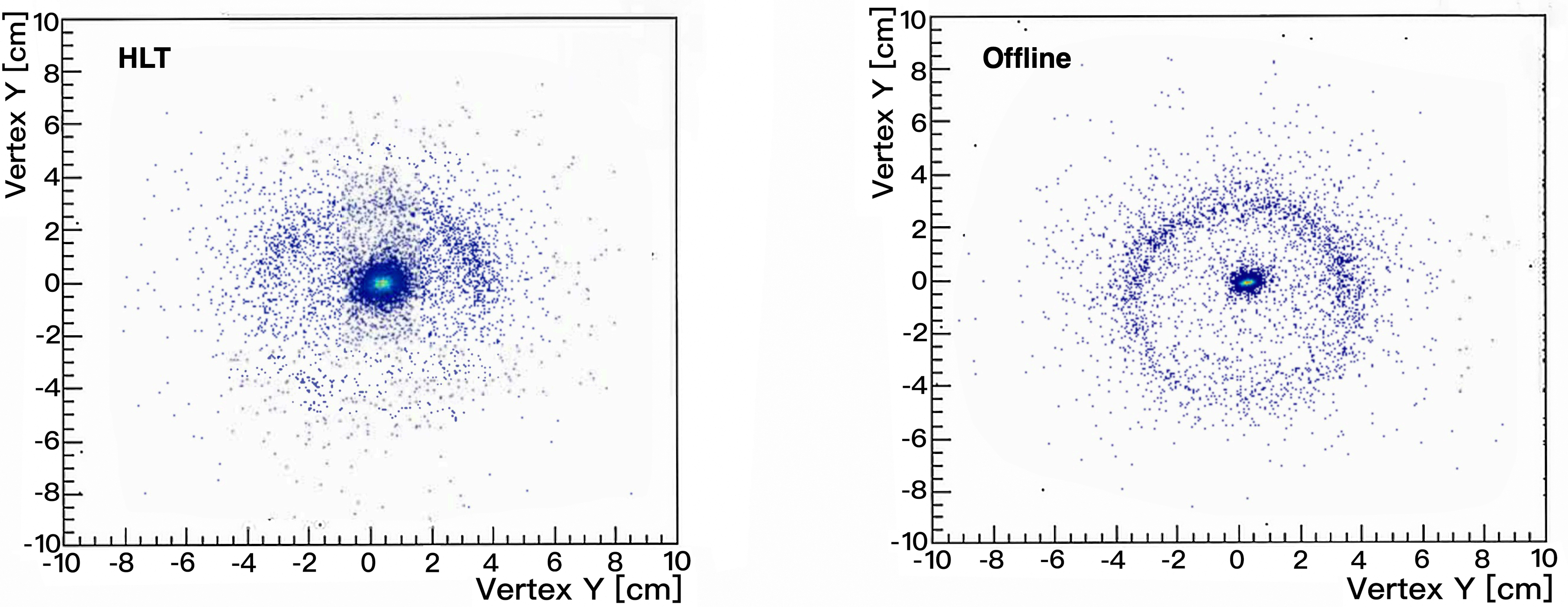}
\caption{\label{fig:Vxy_L4_offLine}
The vertex positions in the x and y coordinates for events reconstructed by HLT (left) and offline (right) are shown. The ring structure is a result of collisions between the beam and the beam pipe.}
\end{figure}

\begin{figure}[htbp]
\centering
\includegraphics[width=1.0\textwidth]{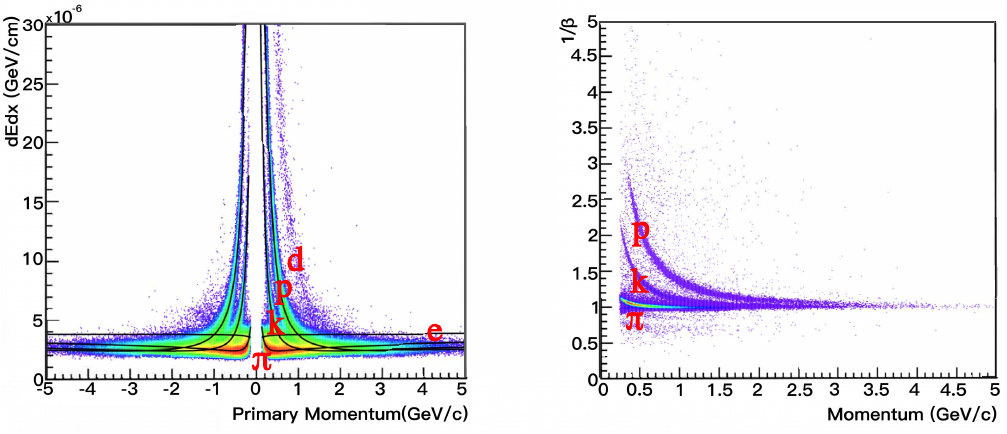}
\caption{Left: The energy loss per unit length as a function of primary momentum for tracks within the TPC, reconstructed by HLT. The primary momentum is the refit momentum with the primary vertex used as an additional hit. Lines are expected band centers for corresponding particle species; antiparticles are shown as mirrored bands.  Right: The $1/\beta$ as a function of momentum for hits measured by TOF that are matched with a TPC track, reconstructed by HLT; particles and antiparticles are overlaid.}
\label{fig:dEdx_P-TOF_inverseBeta_P}
\end{figure}

In Fig.~\ref{fig:Vxy_L4_offLine}, the HLT and offline primary-vertex maps agree
without visible offset or distortion. By design,
offline reconstruction remains the reference for final resolution, while HLT prioritizes
low latency and unbiased, stable estimates adequate for L4 decisions and beam
monitoring. The operational use of the HLT primary vertex as real-time feedback to
the accelerator further corroborates its accuracy for run-time needs. In Fig.~\ref{fig:dEdx_P-TOF_inverseBeta_P}, left, the energy loss per unit length (dE/dx) of track inside the TPC is plotted against the reconstructed momentum p, it can be seen the characteristic band corresponding to various particle species can be clearly identifiable. A similar plot is shown in Fig.~\ref{fig:dEdx_P-TOF_inverseBeta_P}, right, for the $1/\beta$ versus momentum reconstructed by TOF, where $\beta$ is the speed of the charged particle, which can be calculated from the known particle momentum and measured time for each particle mass hypothesis in turn. 

Additionally, the HLT includes a Live Event Display that offers a graphical representation of reconstructed events. This display shows tracks, vertices, and detector hits in a 3D format, presenting complex event information in an accessible way. By monitoring event features and spatial distributions, operators can assess beam quality, detect potential issues early, and initiate prompt troubleshooting. This combination of real-time QA metrics and visual event diagnostics supports efficient data-taking, minimizes downtime, and helps safeguard the quality of data recorded for analysis.

QA outputs are produced run-by-run and reviewed by the shift crew; Plots are generated by
ROOT-based scripts
and published as  on an web page hosted by the HLT control server. If a misalignment or persistent issue is suspected, the shift crew pages the relevant subsystem/HLT experts. Updated constants from the offline database are pushed via the HLT control server and picked up by the tracking nodes within minutes. If needed, L4 thresholds may be temporarily tightened or relaxed, or data taking paused and then resumed after a targeted fix. In practice, such interventions are infrequent; significant retuning of alignment and/or space-charge calibration is typically required only when RHIC running conditions change substantially (e.g., a switch of energy and/or species).

\subsection{Extended Functionality of the High Level Trigger (xHLT)}

\begin{figure}[htbp]
\centering
\includegraphics[width=1.0\textwidth]{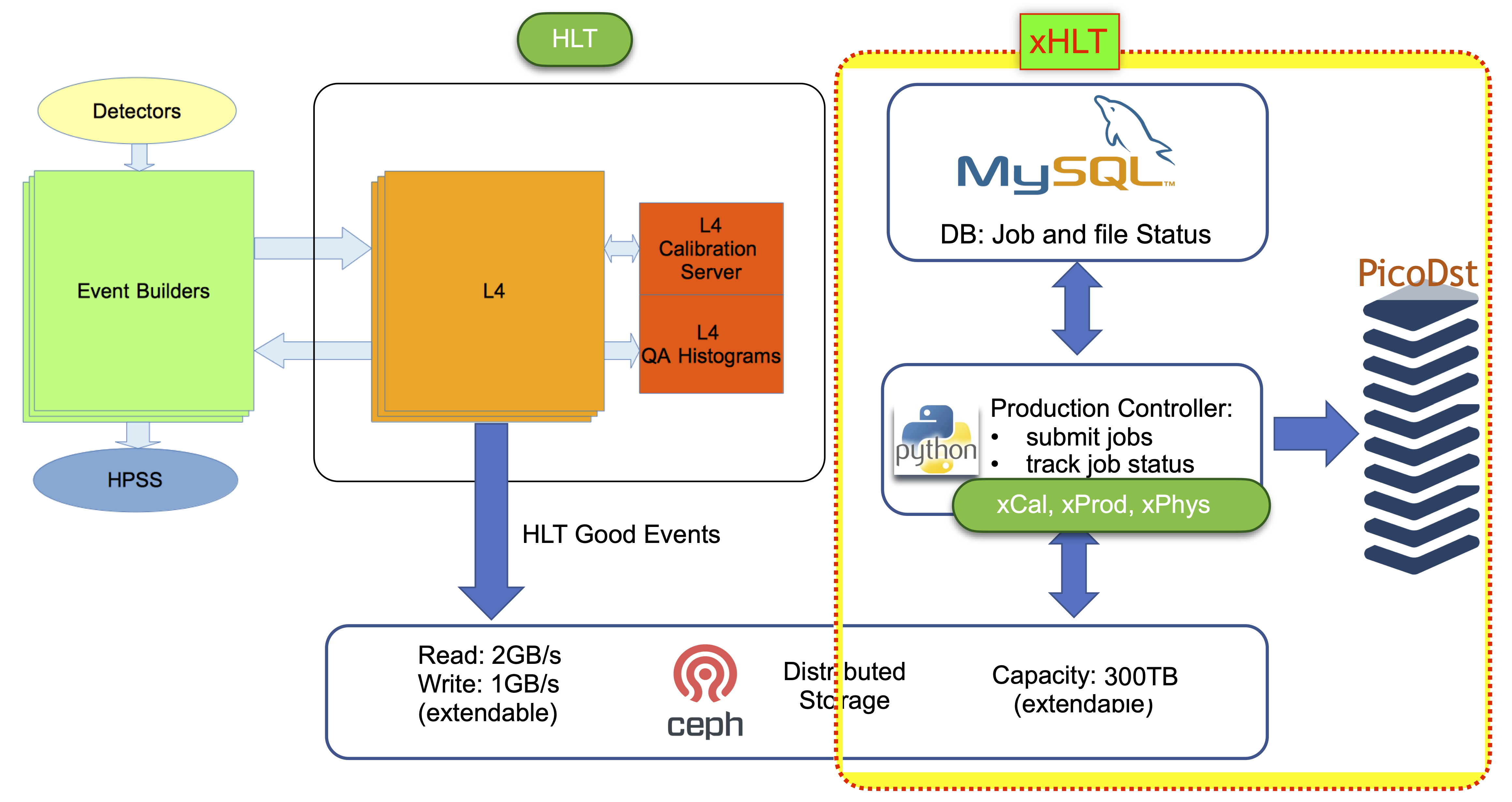}
\caption{Architecture and data flow of HLT and xHLT. 
Left: DAQ integration. Detectors feed the event builders, which deliver events to HLT (L4); the L4 calibration server and online QA histograms run within HLT. HLT-good events are written to the distributed storage, while the full raw stream is logged to HPSS. Right: xHLT services on the HLT cluster. A MySQL database records job and file status; a Python production controller submits and tracks jobs for xCalibration, xProduction, and xPhysics; xProduction converts the buffered HLT-selected DAQ files to compact picoDsts, which are written on HLT storage}

\label{fig-STAR-xHLT} 
\end{figure}

Additional HLT features are detailed in Fig.~\ref{fig-STAR-xHLT} and include advanced filtering,
storage, and data transfer capabilities. The underlying storage system provides
300~TB storage space and supports up to 2~GB/s read and
1~GB/s write bandwidth. This footprint buffers multiple days of HLT-good data
($\sim$25–30~TB per 16~h day at $\sim$450~Hz) and supports concurrent xProduction
reads/writes at DAQ-delivered rates, decoupling xProduction from tape staging
and enabling rapid turnaround. Data accumulate on the HLT disks for immediate access and
low-latency processing. These HLT-selected events are written to the HLT distributed storage in DAQ format and are converted to picoDst by xProduction for physics working groups (PWGs) to access for further
analysis.

Before BES–II, the HLT provided the baseline online chain, including CA-based tracking,
primary-vertex finding, L4 event selections, and run-by-run QA.
The BES–II extension builds on this by integrating xCalibration,  xProduction to picoDst, and xPhysics 
into the existing HLT farm and workflow.
Here, The xCalibration is a nearline snapshot
calibration service that builds offline-format tables from recent DAQ files
buffered on the HLT disks and distributes them within hours. The xProduction is an independent reconstruction chain that runs in parallel on the HLT cluster, and writes compact picoDsts for rapid QA and early
physics checks.  The xPhysics is a lightweight
physics-level selectors and monitors that consume xProduction outputs and/or
HLT tracks to validate signals during running. The detailed implementation is
presented in Sec. ~\ref{sec:RecoAndSoftWare} and ~\ref{sec:xProduction}. To create a package of algorithms for the full processing and analysis of data in real time within the BES-II physics program, the functionality of the HLT computer cluster was significantly extended and integrated into a unified online workflow (xCalibration, xProduction, xPhysics). This was done within the FAIR Phase-0 program, which allowed adapting the package of fast algorithms for processing and analysis FLES (First Level Event Selection)~\cite{IKisel} of the CBM experiment (FAIR/GSI) data to work with real data of the STAR experiment. 

Basic elements of xProduction and xPhysics are search of particle trajectories in the detector system based on the CA Track Finder and search of short-lived particles based on the Kalman Filter (KF) Particle Finder~\cite{IKisel}. These algorithms, after careful adaptation and detailed testing on simulated data, were validated across datasets spanning multiplicity and energy:
p+p (2013), Au+Au (2014, 2016), and BES-II collisions, with p+p serving as a
comparatively low-multiplicity baseline for performance checks and comparisons. It was shown that using the CA Track Finder provides 25\% more D$^0$ and 20\% more W relative to STAR’s then-standard
offline track finder when processing a sample of pp collisions at $\sqrt{s}$ = 510 GeV collected in 2013. Also, the KF Particle Finder package provides~\cite{Ju:2023xvg} twice the background-subtracted signal yield
(invariant-mass peak) at the same combinatorial background level relative
to STAR’s then-standard offline topological reconstruction
(pairwise DCA/pointing-angle selections without a global KF fit).

An important component of the system is its tight integration with data QA mechanisms. Working in real time, the xHLT not only selects and processes data, but also monitors its quality with physics observables, reducing the risk of errors in later stages of analysis. 

These advanced xHLT capabilities support the physical analysis of BES-II data and make the entire HLT system more robust, flexible, and scalable for different real-time environments.

\section{Reconstruction Software}\label{sec:RecoAndSoftWare}

Both the HLT and the xProduction chain depend on a shared set of real-time reconstruction algorithms derived from the STAR offline framework and executed on the HLT computer cluster. These algorithms are designed to sustain event rates up to 2~kHz (with several GB/s throughput) and maintain tracking efficiency above 95\% with near-offline precision. This enables the reliable reconstruction of rare signals during DAQ. Tracking purity is commonly summarized by clone and ghost fractions (ghost defined via track purity); operationally, these contributions are required to remain sub-dominant (at the 10\% level or below before analysis-quality cuts) so they do not bias online QA observables or inflate combinatorial background for secondary-vertex reconstruction. These are discussed for the STAR CA tracker in Ref.~\cite{GKozlov} and are controlled by standard STAR track-quality selections used in both HLT and xProduction. The CA Track Finder~\cite{GKozlov} reconstructs charged particle trajectories by forming hit triplets in the TPC sectors and iteratively linking them into track candidates. This method is optimized for STAR’s 24-sector geometry, SIMDized, and parallelized across CPU cores. The KF Particle Finder~\cite{MZyzak, PKisel} extends this by fitting decay topologies with a Kalman filter and identifying secondary vertices consistent with weak decays or light hypernuclei. Together, these algorithms deliver fast, parallelized reconstruction with high efficiency. They provide online event selection in the HLT and near offline-quality results in  xProduction.

\subsection{Cellular Automaton Track Finder: Search for Long-Lived Particles}\label{subsec:CATrackFinder}

The CA is used in the STAR experiment to reconstruct the trajectories of charged particles produced in heavy ion collisions. The main tracking detector, the TPC, registers charged particles as they pass through its gas volume and generates a sequence of space points (hits) along their paths. Each hit is characterized by three spatial coordinates ($x, y, z$), and the cylindrical geometry of the TPC, segmented into 24 sectors, defines a natural division for tracking. Most particles move outward from the point of interaction toward the periphery of the detector, so it is advantageous to handle hits in local sector-based coordinates.  

Due to the high particle multiplicity (up to 5000 per event) and the complex detector geometry, a sector-wise approach is used for the track finding, similar to the one used in the ALICE experiment~\cite{ALICE:2008ngc}. This method simplifies the combinatorial problem by restricting the initial search to within sectors before merging partial tracks across sector boundaries. Also, local reconstruction allows to improve the accuracy and stability of the method, because hit position together with its errors are defined in the local coordinate system. The position of the pad row sets a fixed hit coordinate (with zero error) along the radial direction of the sector center, that simplifies track propagation compared to the global track model. Since the majority of the particles are flying along the radial direction in the XY plane, operation in local coordinate system helps to minimize numerical issues. 

The CA Track Finder proceeds through several key steps. First, it identifies triplets -- groups of three consecutive hits -- without generating intermediate singlets or doublets, optimizing processing speed. Triplet formation includes competitive selection to ensure geometric consistency based on slope angles. These triplets are then subjected to an evolutionary selection process that retains a triplet only if it shares two hits with another triplet, effectively filtering out spurious candidates. To improve the reconstruction of low-momentum particles, an additional pass of triplet search is performed, excluding previously used hits, thereby capturing soft tracks more efficiently. Triplets are extended into longer chains by propagation to the next pad row, searching hits intersecting with the track, and fitting with a Kalman filter, which allows the construction of tracklets that tolerate a small number (up to 4) of missing hits.  

Finally, tracklets from different sectors are merged into global tracks. This requires precise comparison of track parameters -- such as position, slope and momentum -- at sector boundaries. The merging procedure includes (1)  sorting of track indices by pad rows to reduce the combinatorial search; (2) matching of track parameters, with special handling of adjacent sector tracks using coordinate transformations to resolve local system differences; and (3) merging and re-fitting, where successfully matched segments are combined into a single track and re-optimized for accuracy.  

\begin{figure}[htbp]
\centering
\includegraphics[width=1.0\textwidth]{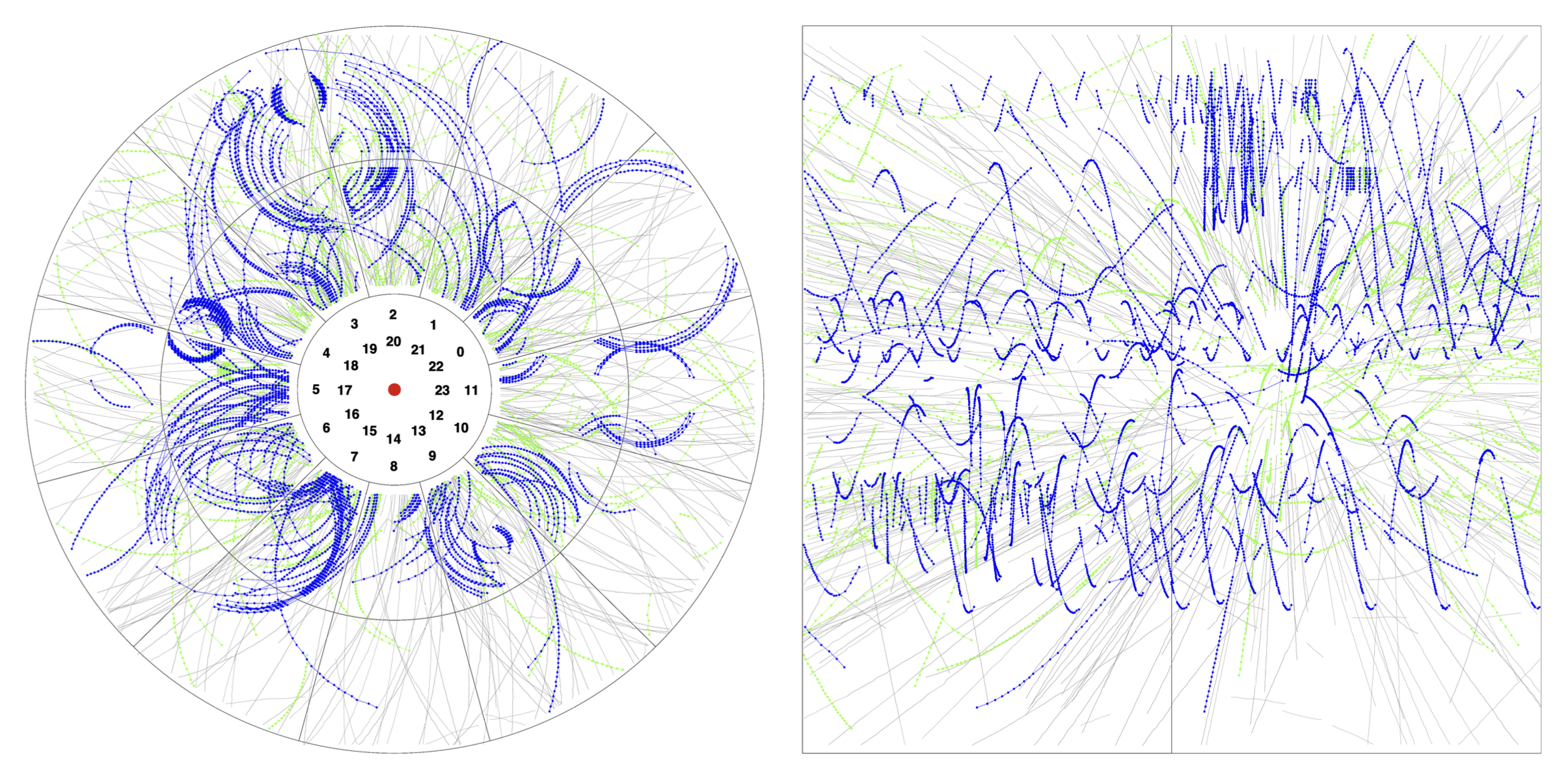}
\caption{An example of a collision containing loopers reconstructed with the CA Track Finder in the TPC detector~\cite{GKozlov}, in end view (left panel) and side view (right panel).  Tracks with a momentum greater than 200 MeV/c are shown in gray, tracks with a momentum less than 200 MeV/c that are not combined into loopers are shown in green, and reconstructed loopers are shown in blue.}
\label{Loopers}
\end{figure}

This tracklet merging system allows all reconstructed segments to be combined into global tracks that represent the final reconstruction result. The robustness and speed of this method allows it to handle large amounts of data and provide high quality trajectory reconstructions for STAR analysis~\cite{GKozlov}. An example of reconstructed tracks, including low-momentum loopers, is shown in Fig.~\ref{Loopers}.

\subsection{Primary Vertex Search: Reconstruction of Collision Point}

The fixed-target mode of the BES program introduces additional challenges for primary vertex reconstruction. At lower beam energies, the beam profile becomes broader, increasing the rate of interactions with the beam pipe material. Furthermore, due to the high interaction rate (up to 2~kHz), several consecutive collisions may overlap within the TPC volume, so in addition to tracks from the triggered collision, there can be significant contamination from pile-up collisions.

\begin{figure}[htbp]
\centering
\includegraphics[width=1.0\textwidth]{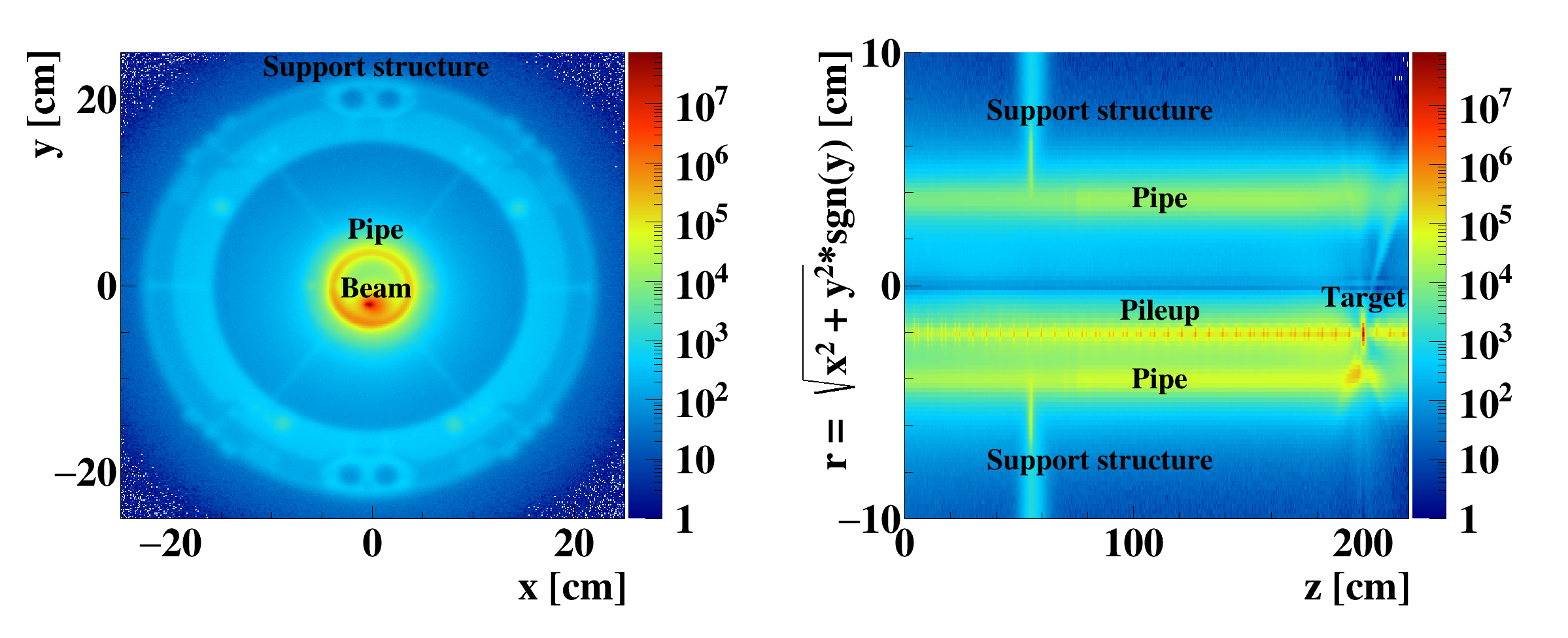}
\caption{Distribution of reconstructed primary vertex position by the multi-vertex reconstruction procedure in Au+Au collisions at $\sqrt{s_{\rm NN}}$ = 3 GeV HLT events from the 2021 run. Left: distribution in the xy plane. Right: distribution in the rz plane, where $r=\sqrt{x^2 + y^2} \cdot \mathrm{sgn}(y)$.}
\label{PV}
\end{figure}

Tracks originating from pile-up and material interactions are typically well separated from those produced in the triggered collision but create substantial background for the reconstruction of short-lived particles, for which daughter tracks are required not to point to the primary vertex. Importantly, since the start time for pile-up tracks is taken from the trigger event, their time reference is incorrect. This leads to a systematic shift in the $z$-coordinate of hits produced by such tracks, as it is determined by drift time in the TPC. Because the time shift is proportional to the collision time difference and to the speed of beam particles (close to the speed of light), all particles from a given pile-up collision will exhibit a similar shift, allowing for their identification and separation. To address this issue, a multi-vertex finder was developed.

In the first step, tracks are clustered according to several hypotheses of their production point (or vertex):
\begin{itemize}
    \item \textbf{Primary vertex}: The main collision point, using tracks near the beam spot at the target position.
    \item \textbf{Pile-up vertices}: Clusters of tracks around the beamline, without restrictions on the $z$-position.
    \item \textbf{Beam pipe interactions}: Clusters formed near the radius of the beam pipe.
    \item \textbf{Detector structure interactions}: Clusters of tracks produced by interactions with detector support structures.
\end{itemize}
\begin{figure}[htbp]
\centering
\includegraphics[width=0.95\textwidth]{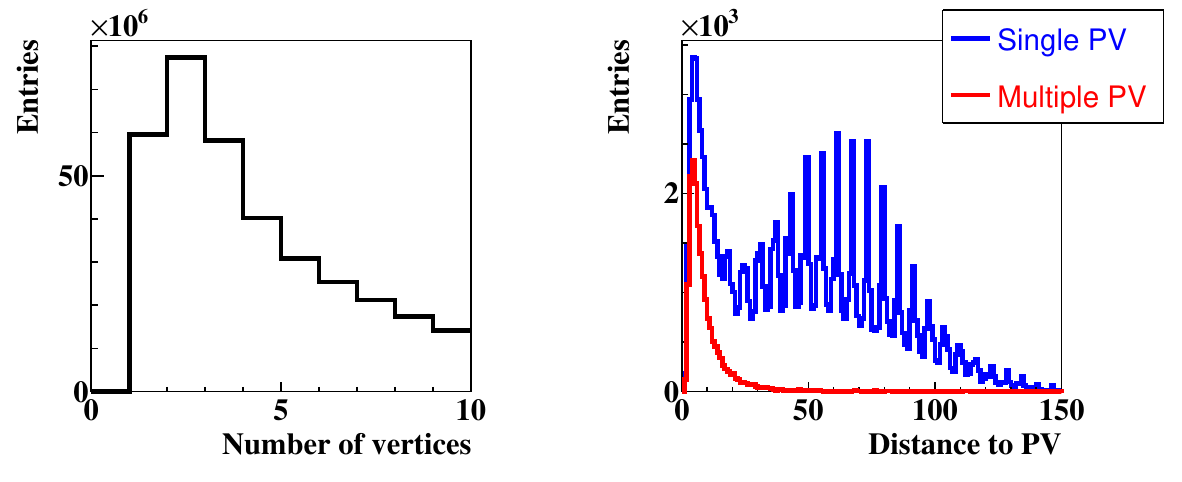}
\caption{Left: distribution of the number of reconstructed primary vertices per triggered event. Right: distribution of distance from reconstructed decay point of ${}^4_{\Lambda}\mathrm{He}\rightarrow^3\mathrm{He} + p + \pi^-$ candidates to the primary vertex using single vertex and multi-vertex reconstruction procedures. Collected with Au+Au collisions at at $\sqrt{s_{\rm NN}}$ = 3 GeV HLT events from the 2021 run.}
\label{NPV}
\end{figure}

The corresponding vertex is then fitted from each track cluster using the Kalman filter~\cite{MZyzak}. The resulting distributions of reconstructed vertices for Au+Au collision at at $\sqrt{s_{\rm NN}}$ = 3~GeV HLT events collected in the 2021 run are shown in Fig.~\ref{PV}, in both the $xy$ and $rz$ planes (where $r = \sqrt{x^2 + y^2} \cdot \mathrm{sgn}(y)$). These distributions clearly reveal key features such as the beam pipe, beam position, and support structures, demonstrating the high quality of detector calibration and alignment. Additionally, the $rz$ projection displays distinct peaks corresponding to pile-up vertices along the beamline, reflecting the bunch structure of the beam.

Due to the high luminosity of fixed-target data taking, each trigger event often contains multiple overlapping collisions. Fig.~\ref{NPV}, left, shows the distribution of the number of reconstructed vertices per event, confirming that multi-vertex events are frequent. Such conditions create significant background for short-lived particle reconstruction, particularly for three-body decays.

The efficiency of the cleaning procedure is illustrated by analyzing the decay $^4_\Lambda\mathrm{He} \rightarrow\ ^3\mathrm{He} + p + \pi^-$. In Fig.~\ref{NPV}, right, the distribution of the distance between the reconstructed decay point of $^4_\Lambda\mathrm{He}$ candidates and the primary vertex is shown, using two methods: assuming a single primary vertex (blue line) and applying the multi-vertex reconstruction (red line). The use of multi-vertex reconstruction substantially reduces the background, improving the reliability of the signal extraction.

\subsection{KF Particle Finder: Search for Short-Lived Particles}\label{subsec:KFParticle}

Short-lived particles, such as hyperons, low-mass vector mesons, and charm particles, decay before reaching the detector and cannot be registered directly. Their properties are studied by reconstructing their decay products -- stable particles like protons and pions, which are tracked by the CA and identified using TPC and TOF detectors. Since particle identification is probabilistic, each track can have several associated hypotheses.  

\begin{figure}[htbp]
\centering
\includegraphics[width=0.98\textwidth]{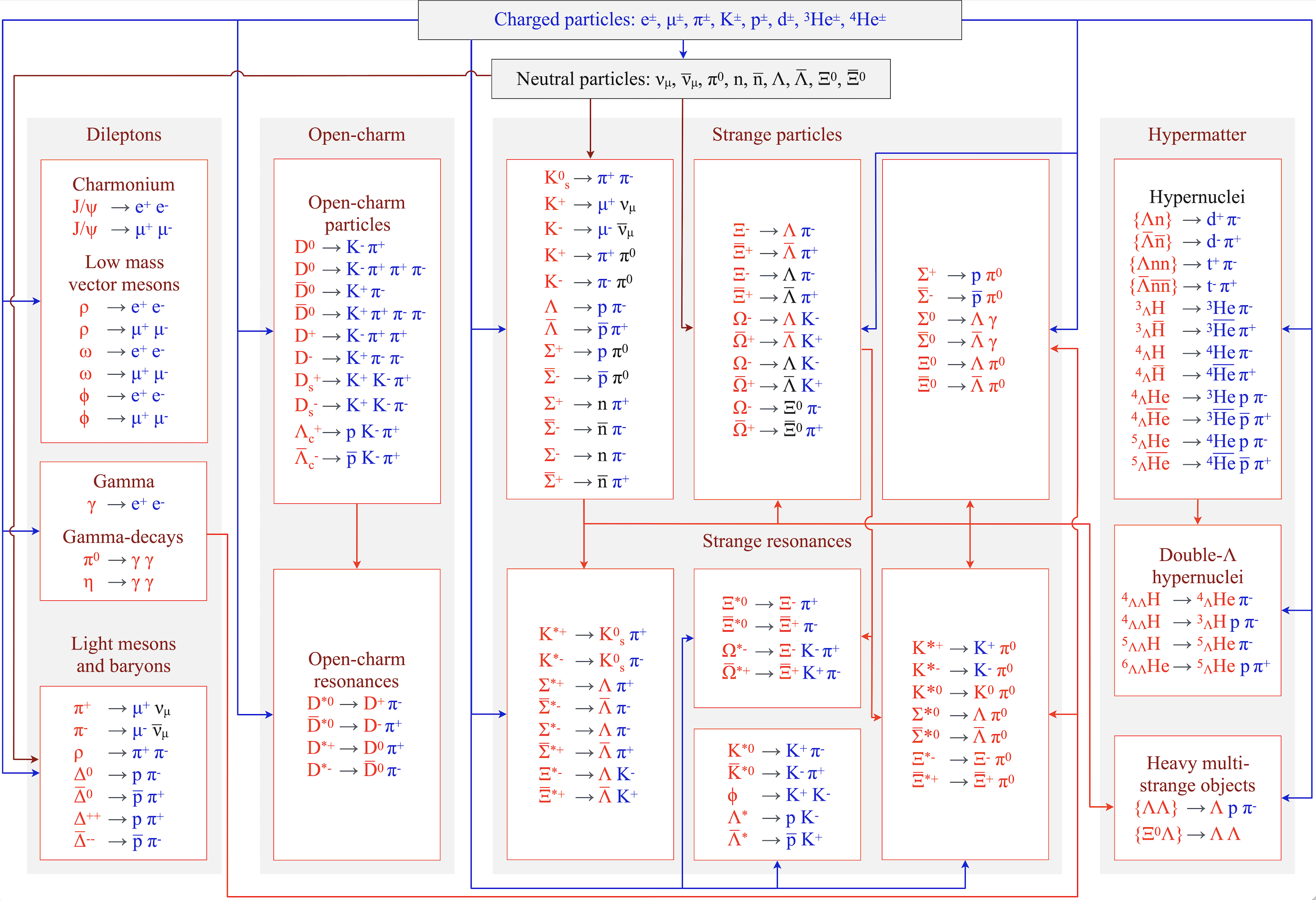}
\caption{A block diagram of the KF Particle Finder package~\cite{PKisel} with the implemented Missing Mass Method for finding decays of short-lived particles with "missing" neutral particles in a daughter channel (highlighted in black).}
\label{KF-Particle-Finder-Block-Diagram-2}
\end{figure}

 Short-lived particles are reconstructed using the KF~Particle and KF~Particle Finder packages~\cite{MZyzak}. Based on the Kalman filter, these tools describe all particles by their position, momentum, energy, and decay time, and allow reconstruction of decay chains with arbitrary complexity. This approach is versatile and applicable to a wide range of experiments, including the search for hyperons and hypernuclei. The KF~Particle Finder extends this framework to real-time identification of short-lived particles, crucial for analyzing rare signals in heavy-ion collisions.  

Reconstruction starts by classifying stable particles into primary (originating from the primary vertex) and secondary (from decays). Secondary particles are separated from background primaries based on their distance to the primary vertex and decay topology. By applying kinematic constraints, the KF~Particle Finder estimates if a set of daughter particles can originate from a common parent, forming a decay candidate. If all topological criteria are satisfied, parameters of the candidate are stored for further analysis.  

Additionally, the KF~Particle Finder implements the Missing Mass Method~\cite{PKisel,A1:2025mjf}, which reconstructs decays with undetected neutral particles by using energy and momentum conservation. For example, in the decay $\Sigma^- \rightarrow n + \pi^-$, where the neutron is not detected, the method uses the pion and decay hypothesis to reconstruct the $\Sigma^-$ mass. Fig.~\ref{KF-Particle-Finder-Block-Diagram-2} illustrates the variety of decay channels analyzed using this approach, including hyperons, open charm, and hypernuclei.  

This methodology enables high-precision reconstruction of short-lived particles, even in high-multiplicity environments, and provides a broad platform for studying rare processes at high interaction rates.

\section{The Express Data Production}\label{sec:xProduction}

The STAR experiment has provided a perfect machinery for studying strange matter for more than two decades~\cite{STAR:2006nmo,STAR:2019bjj}.  Building on this program, an express procedure was developed, enabling online monitoring of the collected physics data. The high quality of express calibration and reconstruction provides a unique possibility to run the express production and observe almost in real-time strange particles including mesons, hyperons, resonances and even hypernuclei. 

\begin{figure}[htbp]
\centering
\includegraphics[width=0.95\textwidth]{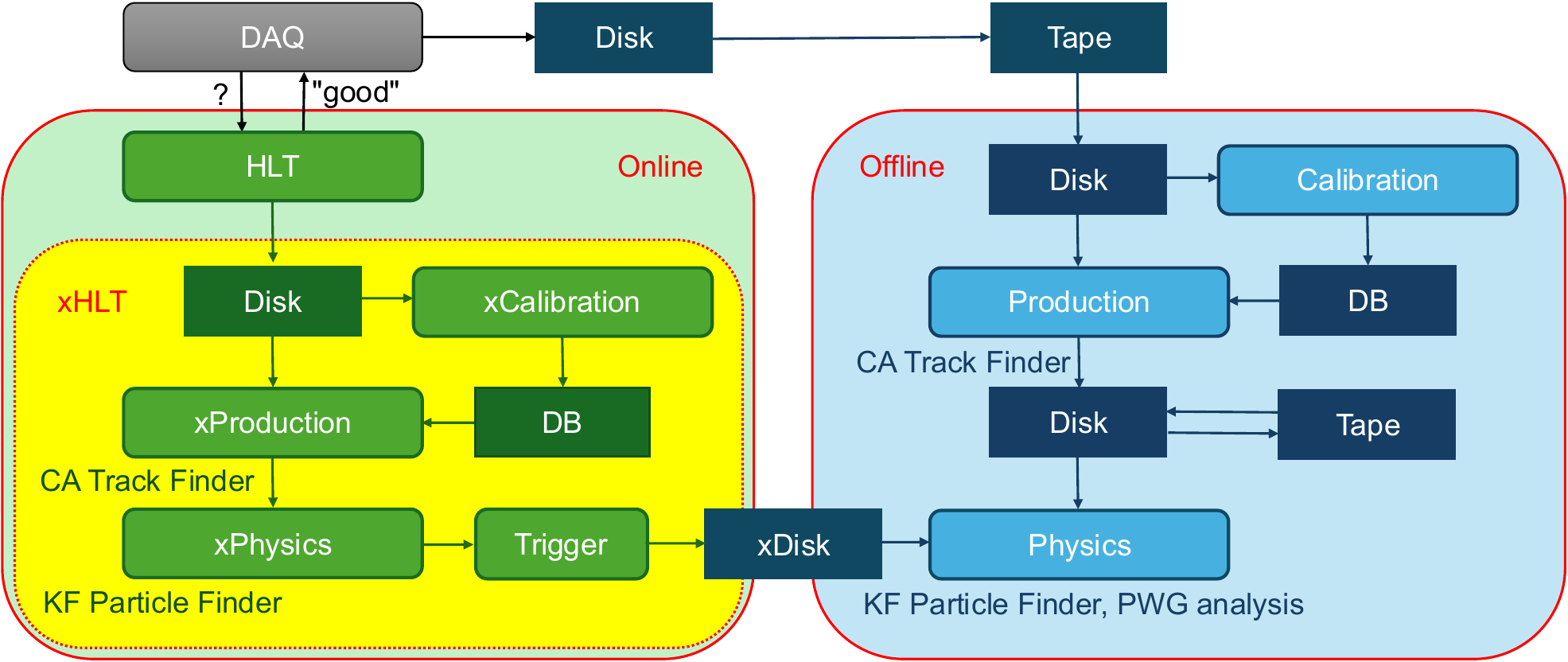}
\caption{BES-II: eXpress+Standard Data Production and Analysis.}
\label{fig-Express-chain} 
\end{figure}

In order to analyze the newly collected data in near real-time and to provide an early access to them for PWGs, a chain of express production and analysis was proposed. As in case of offline, such chain consists of alignment and calibration (xCalibration), reconstruction of hits and particle trajectories (xProduction), and reconstruction of the primary vertices and short-lived particles together with their analysis (xPhysics). Express (online) processing uses the same algorithms as the offline, ensuring that preliminary and final physics results are compatible and reducing the need for recalculations or adjustments. This unified solution helps to achieve reliable results in the shortest possible time, making the system more flexible and adaptable. Using the same algorithms, the proposed scheme, however, does not affect the standard offline reconstruction to maintain a consistent approach to data processing at all stages.

The prerequisites were, that the express production should not overlap with already existing HLT tasks and should not interfere with the well established STAR DAQ system, i.\ e.\ should run in parallel with it and should not compete with the standard HLT routines for the compute resources. Therefore the complete express production has been running exclusively at the idle part of the HLT computer farm. The HLT compute resources are sufficient to process the data at about 0.3~kHz interaction rate. In runs 2019 and 2020, all data were processed, given the available resources. In the 2021 fixed-target run the interaction rate was increased up to 2~kHz. In order to enrich the processed sample with collisions of the particular physics interest we introduced a software trigger.

The xHLT chain added on top of the standard HLT processing is illustrated in Fig.~\ref{fig-Express-chain} and highlighted with a yellow box. As follows from the design of xHLT, the reconstruction scheme is similar to the offline. If reconstructed events are passing the software trigger, they are stored on the disk in the standard STAR data format directly suited for the physics analysis. Therefore they can be used by PWGs immediately after the data are collected and processed. 

At the energies of the STAR BES-II program~\cite{BES-II} strange particles including strange mesons, hyperons, and hypernuclei are one of the main observables. Light hypernuclei are expected to be abundantly produced in low energy heavy-ion collisions, and the BES-II program provides a unique opportunity to study their properties~\cite{Chen:2024aom}. Therefore, with a focus on hypernuclei, a helium trigger was introduced, requiring the presence of a He nucleus in the event, for subsequent real-time analysis. A set of 437 M triggered Au+Au collisions at $\sqrt{s_{\rm NN}}$~=~3.0~GeV in the fixed-target mode recorded in 2021 on HLT proved to be sufficient to measure the yield, lifetime, and spectra of the hypernuclei. In addition, strange particles and hyperons serve as optimal candidates for data quality monitoring through express stream analysis.

\subsection{BES-II: Search for Mesons and Hyperons}

The quality of the express chain of data processing and analysis was continuously tested in real time using meson reconstructions as an example. For reconstruction of short-lived particles, the KF~Particle~Finder package was used with its standard reconstruction scheme, including the standard list of selection criteria. PID of daughter tracks was performed using HLT algorithms discussed above.

\begin{table}[htbp]
\centering 
\caption{Results of the search for mesons and hyperons in the STAR experiment HLT express data stream for Au+Au collisions at $\sqrt{s_{\rm NN}}$~=~7.7~GeV. The top and middle blocks use 32.5 M events; the hyperon block uses 140 M events.}
\label{tab-mesons-hyperons}  
\begin{tabular}{lccccc}
\hline
Decay
  & Mass            & $\sigma$        & $S$               & $S/B$      & $S/\sqrt{S+B}$ \\
  & (MeV/$c^2$)     & (MeV/$c^2$)     &                   &            &                \\
\hline
$\pi^0 \rightarrow \gamma_{e^+e^-} \gamma_{e^+e^-}$ & ~~135.9         &     4.2             & 14.0 $\cdot 10^3$  &       ~~0.2 &     ~~~48       \\
$K^0_s \rightarrow \pi^+\pi^-$                      & ~~497.3         &     4.1             & 67.1 $\cdot 10^6$  &       ~~6.5 &      7629       \\
$K^+ \rightarrow \pi^+\pi^+\pi^-$                   & ~~493.9         &     2.6             & ~~2.4 $\cdot 10^6$ &       24.3  &      1524       \\
$K^- \rightarrow \pi^+\pi^-\pi^-$                   & ~~493.9         &     2.4             & ~~0.7 $\cdot 10^6$ &       ~~8.3 &     ~~839       \\
$K^+ \rightarrow \pi^+\pi^+\pi^-$ + $K$ track       & ~~493.8         &     2.0             & 35.7 $\cdot 10^3$  &       n/a   &     ~~189       \\
$K^- \rightarrow \pi^+\pi^-\pi^-$ + $K$ track       & ~~493.8         &     2.0             & 12.9 $\cdot 10^3$  &       n/a   &     ~~114       \\\hline
$\pi^+ \rightarrow \mu^+\nu_\mu$                    & ~~138.2         &     2.2             & ~~2.1 $\cdot 10^6$ &       75.9  &      1443       \\
$\pi^- \rightarrow \mu^-\bar{\nu}_\mu$              & ~~138.2         &     2.2             & ~~2.4 $\cdot 10^6$ &       78.6  &      1546       \\
$K^+ \rightarrow \mu^+\nu_\mu$                      & ~~493.8         &     9.1             & ~~3.1 $\cdot 10^6$ &       ~~4.7 &      1606       \\
$K^- \rightarrow \mu^-\bar{\nu}_\mu$                & ~~493.6         &     9.0             & ~~1.1 $\cdot 10^6$ &       ~~4.5 &     ~~956       \\
$K^+ \rightarrow \pi^+\pi^0$                        & ~~493.2         &     6.7             & ~~1.0 $\cdot 10^6$ &       ~~2.6 &     ~~830       \\
$K^- \rightarrow \pi^-\pi^0$                        & ~~493.1         &     6.6             & ~~0.3 $\cdot 10^6$ &       ~~2.4 &     ~~489       \\\hline
$\Lambda \rightarrow p\pi^-$                        & 1115.7          &     1.5             & 60.1 $\cdot 10^6$  &       24.6  &      7601       \\
$\bar{\Lambda} \rightarrow \bar{p}\pi^+$            & 1115.7          &     1.4             & ~~0.9 $\cdot 10^6$ &       ~~7.1 &     ~~931       \\
$\Xi^- \rightarrow \Lambda\pi^-$                    & 1321.9          &     2.1             & ~~0.8 $\cdot 10^6$ &       21.8  &     ~~890       \\
$\bar{\Xi}^+ \rightarrow \bar{\Lambda}\pi^+$        & 1321.9          &     2.1             & 45.8 $\cdot 10^3$  &       36.7  &     ~~211       \\
$\Omega^- \rightarrow \Lambda K^-$                  & 1672.4          &     2.2             & ~~9.2 $\cdot 10^3$ &       ~~3.9 &    ~~~~86       \\
$\bar{\Omega}^+ \rightarrow \bar{\Lambda} K^+$      & 1672.4          &     2.4             & ~~2.2 $\cdot 10^3$ &       12.0  &    ~~~~46       \\\hline
\end{tabular}
\end{table}

The upper part of Tab.~\ref{tab-mesons-hyperons} shows the results of reconstruction of decay channels 
$\pi^0 \rightarrow \gamma_{e^+e^-} \gamma_{e^+e^-}$, $K^0_s \rightarrow \pi^+\pi^-$, $K^+ \rightarrow \pi^+\pi^+\pi^-$, and $K^- \rightarrow \pi^+\pi^-\pi^-$ 
after processing 140 M Au+Au events at $\sqrt{s_{\rm NN}}$ = 7.7 GeV, collected in 2021. Due to the high quality of online calibration and processing, strange mesons are reconstructed with high significance and S/B ratio, and even $\pi^0$ is observed with a significance of $48\sigma$. Reconstruction of $\pi^0$ relies on a rather complex search of photons, as electron and positron are parallel at the conversion point, and requires high efficiency of track finding, since 4 tracks are produced in the decay tree.

\begin{figure}[htbp]
\centering
\includegraphics[width=1.0\textwidth]{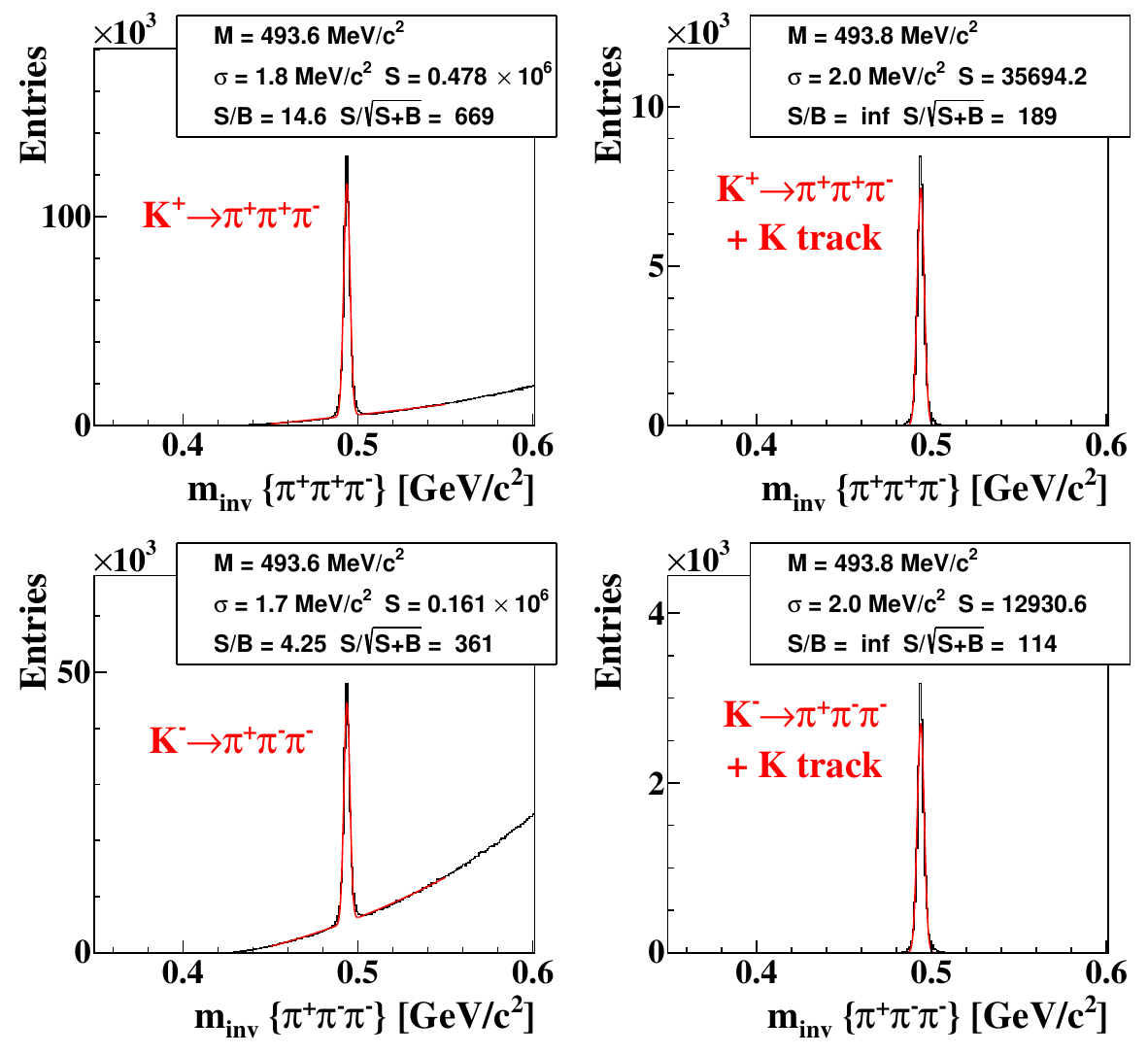}
\caption{Charged kaons reconstructed in 32.5 M Au+Au events at $\sqrt{s_{\rm NN}}$ = 7.7 GeV from the 2021 express data stream using. Left side --- reconstructed using only 3 pion tracks, right side --- all 4 tracks including kaon.}
\label{fig-BES-II-k3pi}
\end{figure}

Also, there is an example of processing 32.5M Au+Au events at $\sqrt{s_{\rm NN}}$ = 7.7 GeV to search for decays $K^+ \rightarrow \pi^+\pi^+\pi^-$ and $K^- \rightarrow \pi^+\pi^-\pi^-$ when all four tracks are registered in the detector system and reconstructed. STAR with its  TPC detector allows to identify charged kaons without background by full topological reconstruction with all 4 tracks including kaon. Reconstruction of such full decay topologies provides additional technical opportunities to study quality of detector performance and reconstruction algorithms. Fig.~\ref{fig-BES-II-k3pi} illustrates both cases: 1) when the decay is reconstructed using pion tracks only and 2) when the kaon track is also used in the reconstruction.

Decay channels of pions and kaons with a neutral daughter particle can also be found using the missing mass method. The middle part of Tab.~\ref{tab-mesons-hyperons} shows the results of the reconstruction of decay channels $\pi^+ \rightarrow \mu^+\nu_\mu$, $\pi^- \rightarrow \mu^-\bar{\nu}_\mu$, $K^+ \rightarrow \mu^+\nu_\mu$, $K^- \rightarrow \mu^-\bar{\nu}_\mu$, $K^+ \rightarrow \pi^+\pi^0$, and $K^- \rightarrow \pi^-\pi^0$ after processing 32.5 M Au+Au events at $\sqrt{s_{\rm NN}}$ = 7.7 GeV, collected in 2021. 

\begin{figure}[htbp]
\centering
\includegraphics[width=1.0\textwidth]{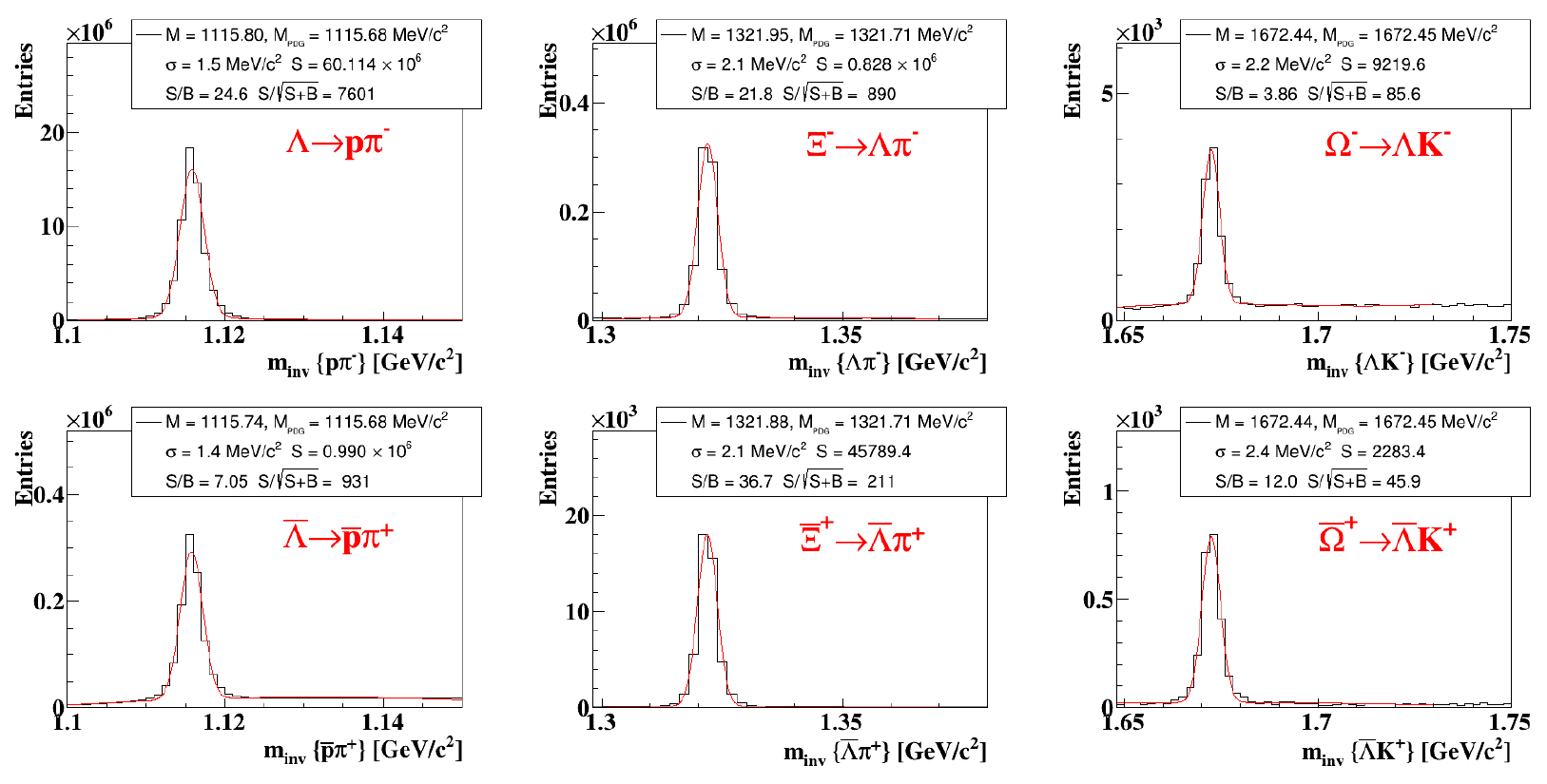}
\caption{Hyperons reconstructed in 140 M Au+Au events at $\sqrt{s_{\rm NN}}$ = 7.7 GeV from the 2021 express data stream using express alignment and calibration.}
\label{fig-BES-II-xHyperons}
\end{figure}

STAR has upgraded the inner part of the TPC in 2019~\cite{STAR:iTPC}, which, together with the enhanced CA track finder, has improved the efficiency of hyperon reconstruction. The bottom part of Tab.~\ref{tab-mesons-hyperons} and Fig.~\ref{fig-BES-II-xHyperons} show the results of reconstruction of the hyperon decay channels after processing 140 M Au+Au events at $\sqrt{s_{\rm NN}}$ = 7.7 GeV, collected in 2021. It can be seen that the high quality of the new BES-II experimental data provides an excellent opportunity to study hyperons. 

\subsection{BES-II: Search for Hypernuclei}

Hypernuclei, nuclei containing one or more hyperons, offer a unique laboratory for exploring the properties of matter under extreme conditions. These exotic systems provide valuable insights into the strong force and the behavior of hyperons, particles that play a crucial role in the structure of neutron stars~\cite{Gal:2016boi,Chen:2025eeb}.

The STAR BES-II program, focusing on the energy range of $\sqrt{s_{\rm NN}}$ = 3.0-27.0 GeV, is particularly well-suited for studying hypernuclei. In this regime, the yields of light hypernuclei (A = 3-5) are expected to reach their maximum at high baryon densities. By analyzing the production and properties of hypernuclei in these collisions, a deeper understanding of the exotic phase of matter that may exist in neutron star cores can be gained.

\begin{figure}[htbp]
\centering
\includegraphics[width=1.0\textwidth]{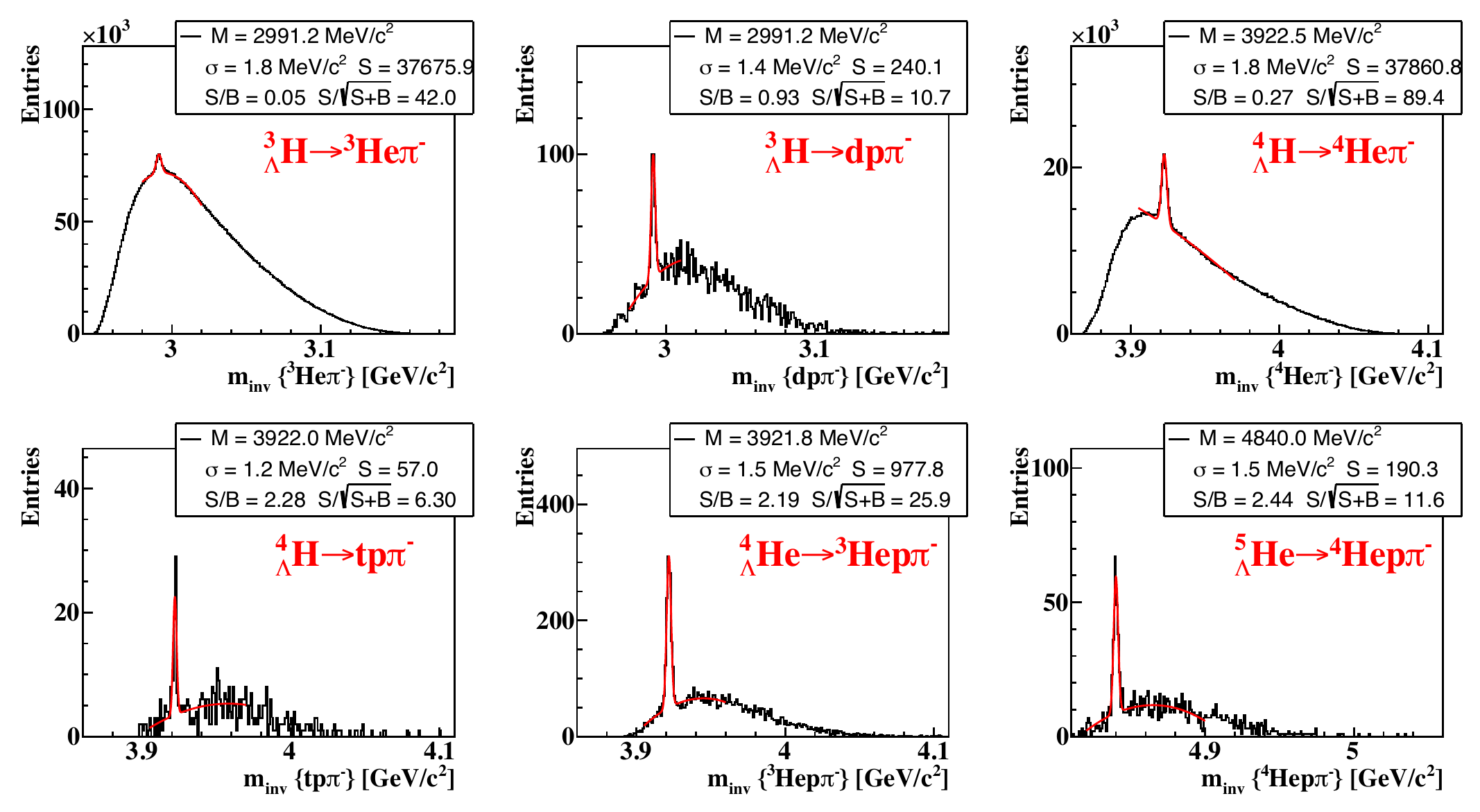}
\caption{Hypernuclei reconstruction using disjoint datasets from 2018–2021 running: the 2021 samples were reconstructed in the express chain, while the 2018–2020 fixed-target samples were reconstructed with the STAR standard production chain after final calibration.} The signal of $_\Lambda^5$He is visible with a significance of 11.6$\sigma$.
\label{fig-BES-II-Hypernuclei} 
\end{figure} 

\begin{figure}[htbp]
\centering
\includegraphics[width=0.99\textwidth]{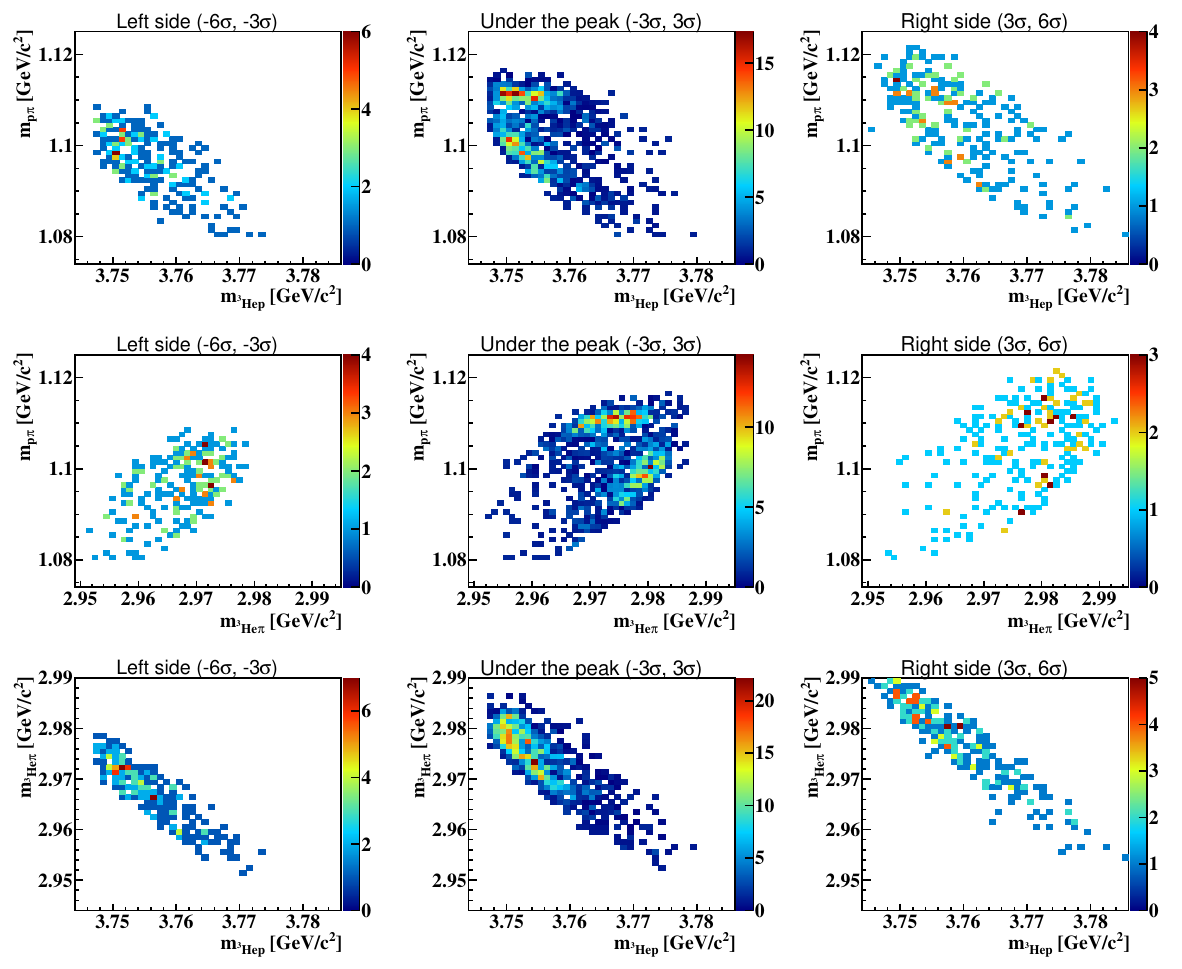}
\caption{The Dalitz plots for the decay $_\Lambda^4$He $\rightarrow$ $^3$He+p+$\pi^-$ which has the largest number of signals (978 decays) found using 2018-2021 data collected at different energies in the collider and fixed-target modes.}
\label{fig-BES-II-Dalitz-1} 
\end{figure}

A total of 437 M triggered Au+Au collisions at $\sqrt{s_{\rm NN}}$ = 3.0~GeV in fixed-target mode, recorded on HLT in 2021, were analyzed. Using the same procedure, the Au+Au collisions at $\sqrt{s_{\rm NN}}$ = 7.7 GeV data in the collider mode collected in 2021 were also analyzed within the express chain, as well as data sets of different energies in the fixed-target mode collected in 2018, 2019, and 2020 and processed with the STAR standard production chain after the final calibration. After such (express and standard) processing of all the data, using disjoint samples from different running periods, the signal of the $_\Lambda ^5$He hypernucleus is clearly visible with a significance of 11.6$\sigma$ (see Fig.~\ref{fig-BES-II-Hypernuclei}). The express-vs-standard reconstruction performance is compared in Table~\ref{tab-hypernuclei}. The $_\Lambda ^5$He hypernucleus is highlighted because it represents the first observation in heavy-ion collisions, and because its multi-body weak decay with a heavy fragment provides a stringent benchmark for tracking/PID and secondary-vertex reconstruction in the xProduction chain.

Collected statistics are also sufficient to study Dalitz plots in the 3-body decay channels. Thus, Fig.~\ref{fig-BES-II-Dalitz-1} shows the Dalitz plots for the decay $_\Lambda^4$He $\rightarrow$ $^3$He+p+$\pi^-$ which has the largest number of signals (978 decays) found. The background was estimated using the side-band method and subtracted under the peak. The complex structure can be explained as a possible spin effect~\cite{STAR:2022zrf} in quasi two-body decays. Similar behavior is observed in the Dalitz plots for the decays of hypernuclei $_\Lambda^5$He $\rightarrow$ $^4$He+p+$\pi^-$ and $_\Lambda^3$He $\rightarrow$ d+p+$\pi^-$, but with lower statistics, 190 and 240 signal particles, respectively.

\begin{table}[htbp]
\centering 
\caption{Reconstruction quality comparison of the express data stream and offline production, where $\sigma$ denotes the fitted invariant-mass peak width from the signal fit. }
\label{tab-hypernuclei}  
\begin{tabular}{lcccccc}
\hline
Decay
  & Enrichment 
  & Trigger
  & Mass
  & Mass diff.
  & $\frac{\sigma_{\mathrm{expr.}}}{\sigma_{\mathrm{offl.}}}$
  & $\frac{S/B_{\mathrm{expr.}}}{S/B_{\mathrm{offl.}}}$ \\
  & factor
  & efficiency
  & (MeV/$c^2$)
  & (MeV/$c^2$)
  & 
  & \\
\hline
$_\Lambda^3$H$\rightarrow ^3$He$\pi^-$   & 1.97              & 0.41         & 2991.2    & 0.1              &   1.00                           & 1.0                         \\
$_\Lambda^4$H$\rightarrow ^4$He$\pi^-$   & 1.64              & 0.34         & 3922.4    & 0.0              &   1.06                           & 1.0                         \\
$_\Lambda^4$He$\rightarrow ^3$Hep$\pi^-$ & 1.30              & 0.27         & 3921.8    & 0.3              &   0.93                           & 0.9                         \\
$_\Lambda^5$He$\rightarrow ^4$Hep$\pi^-$ & 1.30              & 0.27         & 4839.9    & 0.0              &   1.00                           & 1.0                         \\\hline
\end{tabular}
\end{table}

In order to check the quality of the alignment, calibration, and reconstruction algorithms in the express data stream, the same reconstruction chain was applied to the officially produced (offline) 2.11 B Au+Au collisions at $\sqrt{s_{NN}}$~=~3.0~GeV from the 2021 run with fine-tuned calibration and alignment. The comparison at the most challenging hypernuclei case is shown in Tab.~\ref{tab-hypernuclei}.

Due to the limitations of the HLT compute resources and the speed of the DAQ system, not all events could be stored in the express data stream. Therefore a heavy fragment trigger was introduced, which set higher priority for the events containing $^3$He or heavier fragments. This enriched the express sample for hypernuclei, which are among the most challenging decay channels in low-energy fixed-target running. With the available HLT resources,  437 M collisions were processed and stored.

By comparing the number of reconstructed hypernuclei candidates in the express and offline data, the trigger efficiency can be estimated; it varies from 27\% to 41\%. Due to the introduced trigger the express data stream was significantly enriched with hypernuclei. It contains about 30\% more $_\Lambda^4$He and $_\Lambda^5$He, 64\% more $_\Lambda^4$H and two times more $_\Lambda^3$H per event than the complete data set.

The peak position, width of the mass peaks and signal to background ratio in the express and offline data are almost identical, the slight difference is mainly due to the different statistics. This demonstrates the high quality of the HLT alignment, calibration, and reconstruction algorithms. The express data stream provided an early access to the hypernuclei enriched high quality data sample for the physics analysis.

\section{Conclusion}

The HLT in the STAR experiment plays an essential role. It covers a wide variety of tasks from  online monitoring of the detector data quality to early analysis and prediction of the observed physics effects, like observation of antimatter and rare hypernuclei probes.

The algorithms of HLT were gradually developed towards the full event reconstruction including such stages as calibration, reconstruction of hits, tracks, primary vertices, and short-lived particles, as well as express physics analysis.

During processing and analysis of express data stream on the STAR HLT within the BES-II program in 2018--2021 the reconstruction of charged particle trajectories was done in real time by the track finder based on the Cellular Automaton, and the search for short-lived particles and hypernuclei by the KF Particle Finder package based on the Kalman filter.
  
The reconstruction chain of HLT demonstrates high quality similar to the offline procedures. This allows monitoring of the collected data in near live time. For instance, during the 2021 run we demonstrated that fixed target data can be safely collected with the increased collision interaction rate keeping the signal to background ratio and significance high for analysis of strange particles and hypernuclei. 

The express data provides an advantage of early access for the physics analysis. High quality of the collected experimental data and online calibration, and reliable performance of data processing and analysis algorithms enabled the observation and investigation of various hypernuclei up to $_\Lambda^5$He with a significance of 11.6$\sigma$. The Dalitz plots of three-body decays of hypernuclei show complex structures with the possible presence of spin effects. There are also hints that a significant fraction of such three-body decays happen via nuclear resonances.

\section*{Acknowledgment}

The authors express their gratitude to the entire STAR Collaboration. In addition, the authors thank the RHIC Operations Group, and SCDF at RHIC for their support. This work was supported in part by the Office of Nuclear Physics within the U.S. DOE Office of Science, the U.S. National Science Foundation, the National Natural Science Foundation of China, the Federal Ministry of Education and Research (BMBF) of Germany, and the Helmholtz Research Academy Hesse for FAIR, Germany.

\vspace{1cm}

During the preparation of this work we used OpenAI’s chatGPT in order
to streamline some of the phrasing. After using this tool, we reviewed and
edited the content as needed and take full responsibility for the content of the publication.

\bibliography{ref}

\end{document}